\documentclass[journal]{IEEEtran}

\usepackage{amsmath,amssymb}
\usepackage{graphicx}
\usepackage{booktabs}
\usepackage{siunitx}
\usepackage{tikz}
\usetikzlibrary{positioning,arrows.meta,fit,calc,backgrounds,shapes.geometric}
\usepackage{pgfplots}
\pgfplotsset{compat=1.18}
\usepackage{algorithm}
\usepackage{algpseudocode}
\usepackage{xcolor}
\usepackage{url}
\usepackage[hidelinks]{hyperref}

\begin{document}

\title{Physics-Direct FPGA Tooth-Contact Computation\\
for Deterministic Gear Digital Twins}

\author{Jiacheng Miao%
  \thanks{Manuscript prepared \today.}}

\maketitle

\begin{abstract}
Gear digital twins, hardware-in-the-loop rigs, and active vibration
control close a loop around the instantaneous tooth-contact state,
demanding a solver that is real-time, deterministic, and
embeddable---properties that loaded tooth contact analysis (LTCA), a
data-dependent linear complementarity problem (LCP) costing seconds per
mesh phase in fp64, structurally lacks; learned surrogates infer fast but
spend one full LTCA solve per training sample and offer no guarantee
beyond their training envelope. We instead compress the closed-form
contact physics---contact-point localization, principal-curvature
extraction, and the elliptical Hertz solution---directly into a
branch-free fixed-point FPGA datapath, termed \emph{physics-direct},
realized through the fully open-source openXC7 flow (no vendor tools, no
floating-point IP) and validated on retired XC7K480T silicon. Four
SoCs---geometry, Hertz, and two fused paths---pass bit-exact JTAG
readback against a golden model. The on-chip preview tracks LCP body
pressure to within $-22\%$ and loaded transmission error (LTE)
peak-to-peak to $-19\%$, preserving the exact $W^{1/3}$ law; with zero
training it extrapolates in load more accurately than a trained network
($18.9\%$ versus $20.3\%$). Latency is a compile-time constant
($\sigma\!=\!0$ jitter), and the DSP-bound geometry kernel saturates the
fabric near $63$ lanes. Physics-direct is an interpretable, deterministic
alternative to neural surrogacy for embedded tooth-contact estimation.

\end{abstract}

\begin{IEEEkeywords}
Tooth contact analysis, loaded transmission error, FPGA acceleration, fixed-point
arithmetic, gear digital twin, hypoid gears, deterministic real-time, open-source EDA.
\end{IEEEkeywords}

\section{Introduction}
\label{sec:intro}

Gear digital twins, hardware-in-the-loop (HIL) test benches, and active
vibration control all close an estimation or control loop around the
instantaneous tooth-contact state, and therefore demand a tooth-contact
solver that is at once real-time, deterministic, and embeddable. The
reference model---loaded tooth contact analysis (LTCA), in which the
pressure over each engaging tooth pair is resolved by a Boussinesq
half-space influence operator constrained by a linear complementarity
problem (LCP)~\cite{kolivand2009ease,icm2018bevel,litvin2004gear}---is
accurate but heavy and inherently sequential: a single mesh phase costs on
the order of seconds on a CPU, and the LCP is a global, data-dependent
solve that resists fixed-latency execution. This is the wrong shape for an
embedded loop.

The prevailing route to a real-time budget is surrogacy: replace the solver
with a learned model, be it a regression network, a deep transmission-error
predictor, or a physics-informed neural network
(PINN)~\cite{te2023surrogate,ml2025hertzian,pinn2024contact,pinn2025energy,aimbs2025gear}.
Surrogates infer in milliseconds, but at three structural costs.
(i)~Each training sample is itself one full LTCA solve ($\approx\!2$\,s), so
a usable corpus is thousands of solves. (ii)~A fitted network carries no
guarantee outside its training envelope---precisely the off-nominal
operating points a twin must survive. (iii)~It is a black box whose
inference, hosted on a shared CPU or GPU, inherits scheduling jitter.
Meanwhile, FPGA hard-real-time simulation---mature and
deterministic---has been developed almost exclusively for power
electronics and motor drives~\cite{fpga2018igbt,fpga2020latency,fpga2022hil};
to our knowledge no prior work maps hypoid tooth-flank contact geometry
onto an FPGA.

We take the route opposite to surrogacy. Rather than learn the map from
operating point to contact state, we compress the closed-form contact
physics---contact-point localization, principal-curvature extraction, and
the elliptical Hertz solution~\cite{hamrock1977elliptical}---directly into a
fixed-point FPGA datapath; we call this \emph{physics-direct}. The pipeline
is three branch-free kernels: a pose-transform and gap-search stage
(Alg.~\ref{alg:gap}), a baked pseudo-inverse curvature fit
(Alg.~\ref{alg:fit}), and a serialized elliptical-Hertz solver
(Alg.~\ref{alg:bhseq}). Its on-chip output is a single-point \emph{preview}
of the contact ellipse and peak pressure---an engineering-magnitude
estimate that tracks the LCP body pressure to within ${\sim}22\%$ and obeys
the Hertzian $W^{1/3}$ law, not a substitute for full-field LCP at the
tooth-edge peaks that only the LCP resolves. We validate it on retired
data-center silicon (Xilinx XC7K480T, Inspur YPCB-00338) through a fully
open-source flow (openXC7: yosys\,+\,nextpnr-xilinx\,+\,prjxray)~\cite{shah2019yosys,openxc7},
with no vendor tools, no floating-point IP, and JTAG readback against a
golden model.

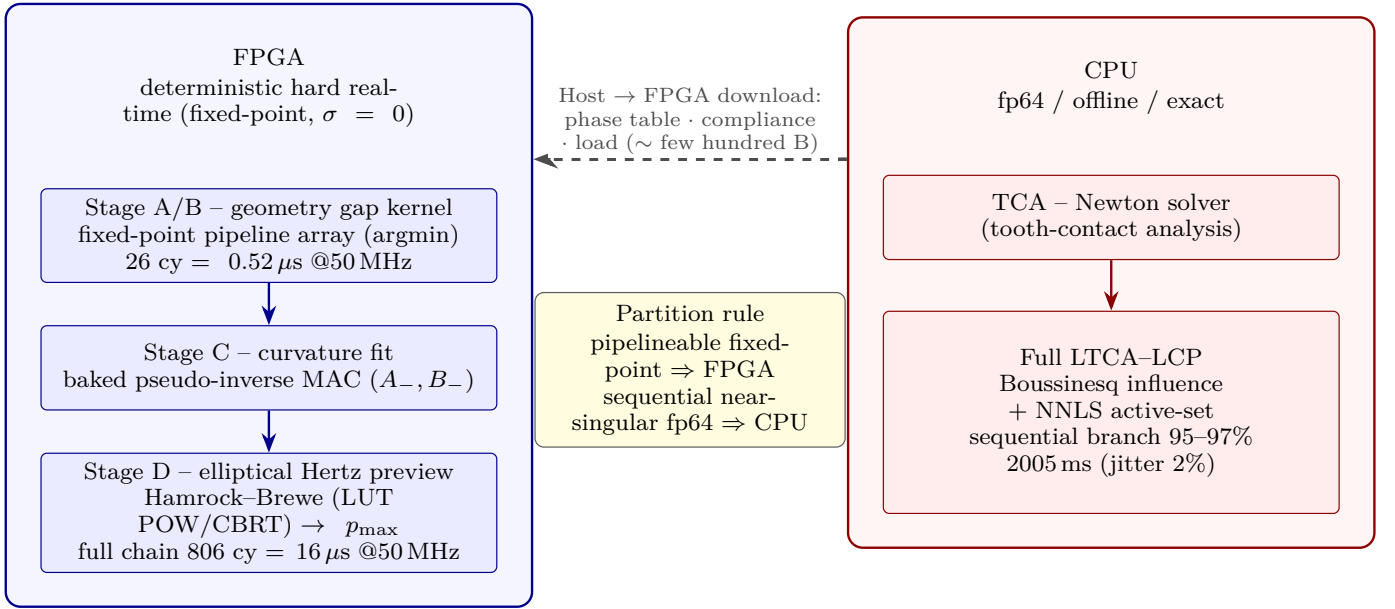
\begin{figure*}[!t]
  \centering
  \resizebox{\linewidth}{!}{%
  \begin{tikzpicture}[
      font=\footnotesize,
      >=Stealth,
      hdr/.style   ={align=center, text width=52mm, font=\footnotesize\bfseries},
      fbox/.style  ={draw=blue!55!black, fill=blue!8,  rounded corners=2pt,
                     align=center, text width=52mm, minimum height=10mm, inner sep=3pt},
      cbox/.style  ={draw=red!55!black,  fill=red!7,   rounded corners=2pt,
                     align=center, text width=52mm, minimum height=10mm, inner sep=3pt},
      lane/.style  ={rounded corners=5pt, inner sep=4mm, thick},
      fflow/.style ={->, thick, blue!55!black},
      cflow/.style ={->, thick, red!55!black},
      dl/.style    ={->, very thick, black!70, dashed},
      crit/.style  ={draw=black!65, fill=yellow!15, rounded corners=3pt,
                     align=center, text width=34mm, inner sep=4pt, font=\footnotesize},
    ]

    \node[hdr] (fhdr) at (0,0)
      {FPGA\\[1pt]\normalfont deterministic hard real-time (fixed-point, $\sigma=0$)};
    \node[hdr] (chdr) at (10,0)
      {CPU\\[1pt]\normalfont fp64 / offline / exact};

    \node[fbox, below=6mm of fhdr] (fa)
      {\textbf{Stage A/B} -- geometry gap kernel\\
       fixed-point pipeline array (argmin)\\
       \emph{26 cy $=0.52\,\mu$s @50\,MHz}};
    \node[fbox, below=5mm of fa] (fb)
      {\textbf{Stage C} -- curvature fit\\
       baked pseudo-inverse MAC ($A_{-},B_{-}$)};
    \node[fbox, below=5mm of fb] (fc)
      {\textbf{Stage D} -- elliptical Hertz preview\\
       Hamrock--Brewe (LUT POW/CBRT) $\to p_{\max}$\\
       \emph{full chain 806 cy $=16\,\mu$s @50\,MHz}};

    \draw[fflow] (fa) -- (fb);
    \draw[fflow] (fb) -- (fc);

    \node[cbox, below=6mm of chdr] (ca)
      {\textbf{TCA} -- Newton solver\\
       (tooth-contact analysis)};
    \node[cbox, below=6mm of ca, minimum height=24mm] (cb)
      {\textbf{Full LTCA--LCP}\\
       Boussinesq influence $+$ NNLS active-set\\
       \emph{sequential branch 95--97\%}\\
       \emph{2005\,ms (jitter 2\%)}};

    \draw[cflow] (ca) -- (cb);

    \begin{pgfonlayer}{background}
      \node[lane, draw=blue!55!black, fill=blue!4, fit=(fhdr)(fa)(fb)(fc)] (flane) {};
      \node[lane, draw=red!55!black,  fill=red!4,  fit=(chdr)(ca)(cb)]     (clane) {};
    \end{pgfonlayer}

    \coordinate (dly) at (0,-0.85);
    \draw[dl] (clane.west|-dly) -- node[above, align=center, font=\scriptsize,
      text width=33mm, fill=white, inner sep=1pt]
      {Host $\to$ FPGA download: phase table $\cdot$ compliance $\cdot$ load
       ($\sim$ few hundred B)} (flane.east|-dly);

    \node[crit] at (5,-3.35)
      {\textbf{Partition rule}\\[1pt]
       pipelineable fixed-point $\Rightarrow$ FPGA\\
       sequential near-singular fp64 $\Rightarrow$ CPU};

  \end{tikzpicture}}
  \caption{CPU/FPGA workload partition, the central thesis of this paper.
  The pipelineable, fixed-point stages of the tooth-contact computation
  (geometry gap kernel, curvature fit, elliptical Hertz preview) map to a
  deterministic FPGA datapath with $\sigma=0$ latency (\num{26} cycles for the
  gap, \num{806} cycles ${=}\,\SI{16}{\micro\second}$ for the full preview at
  \SI{50}{\mega\hertz}),
  whereas the inherently sequential, near-singular fp64 solvers (TCA Newton and
  the LTCA--LCP with Boussinesq influence coefficients and an NNLS active-set,
  whose sequential branch dominates \SIrange{95}{97}{\percent} of the
  \SI{2005}{\milli\second} solve) remain on the CPU. Before each deterministic
  FPGA pass the host preloads only a few hundred bytes (phase table, compliance,
  load).}
  \label{fig:partition}
\end{figure*}

\noindent This paper contributes:
\begin{itemize}
\item \textbf{The first silicon-validated hypoid tooth-contact pipeline on
FPGA.} Four SoCs---the geometry stage, the Hertz stage, and two fused
paths---pass bit-exact JTAG readback against a software golden model on
XC7K480T, the full chain producing relative curvatures, ellipse semi-axes,
and peak pressure on-chip, with cycle-accurate measured schedules
(Fig.~\ref{fig:bhseq}) and bitstream-level die-occupancy evidence
(Fig.~\ref{fig:floorplan}).
\item \textbf{A principled CPU/FPGA partition}
(Fig.~\ref{fig:partition}) that places only the
deterministic, branch-free inner kernels on the fabric while setup and
orchestration remain on the host.
\item \textbf{Physics-direct as a quantified alternative to neural
surrogates.} With zero training, the physics-direct preview extrapolates at
least as well as an MLP fitted to in-range LCP data---$18.9\%$ versus
$20.3\%$ error outside the training range---while remaining interpretable
(every intermediate is a physical quantity) and fixed-point auditable; the
surrogate's accuracy advantage is confined to interpolation
(Fig.~\ref{fig:nn}).
\item \textbf{A floating-point-free, open-source, reproducible realization:}
all arithmetic is fixed-point, and sub-micron gap accuracy needs only
20-bit coordinates (Table~\ref{tab:fixedpoint}), removing any dependence on
fp64 or vendor IP.
\item \textbf{An end-to-end evaluation} spanning accuracy versus load,
loaded transmission error (LTE) as the dominant NVH
excitation~\cite{te2018nvh,evgear2025nvh}, multi-tooth load sharing,
fixed-point precision, resource, throughput, and deterministic latency.
\end{itemize}

The remainder of the paper is organized as follows.
Section~\ref{sec:related} surveys related work and
Section~\ref{sec:background} formalizes the contact problem;
Section~\ref{sec:partition} derives the CPU/FPGA partition and
Section~\ref{sec:pipeline} the fixed-point pipeline;
Section~\ref{sec:impl} covers the open-source implementation and silicon
bring-up; Section~\ref{sec:eval} reports the evaluation; and
Sections~\ref{sec:discussion}--\ref{sec:conclusion} discuss the validity
envelope and conclude.

\section{Related Work}
\label{sec:related}

\emph{TCA/LTCA solvers.}
Tooth contact analysis (TCA) and loaded tooth contact analysis (LTCA) are
the established route from gear geometry to contact pattern and loaded
transmission error (LTE). Litvin's local synthesis and TCA formalism
\cite{litvin2004gear} and Kolivand--Kahraman's ease-off/surface-of-action
LTCA \cite{kolivand2009ease} underpin most modern hypoid and bevel solvers.
Recent work sharpens both speed and fidelity: fast semi-analytical hypoid and
bevel contact \cite{efficient2025hypoid,semianalytical2024bevel}, multi-tooth
loaded models \cite{multitooth2025hypoid}, robust re-formulations of the
meshing equations \cite{robust2024hypoid,nie2024hypoid}, and
influence-coefficient bevel LTCA \cite{icm2018bevel}. All execute offline on
CPU/GPU, coupling dense influence matrices to iterative contact solvers, and
none targets deterministic hardware. We keep their physical basis but
restructure the per-mesh contact computation into a fixed-latency datapath.

\emph{Learning-based surrogates.}
To bypass that cost, data-driven surrogates regress TCA/LTCA outputs:
transmission-error models \cite{te2023surrogate}, Hertzian-pressure regressors
\cite{ml2025hertzian}, and physics-informed networks for contact and strain
energy \cite{pinn2024contact,pinn2025energy,aimbs2025gear}. They evaluate
in microseconds but inherit three liabilities: per-model training sets in which
every label is itself a full solve, opaque interiors, and no error guarantee
outside the training box. We instead evaluate the governing Hertz and curvature
relations directly in hardware---a physics-direct path requiring no training.
Against a full linear complementarity problem (LCP) reference under load
extrapolation, this zero-training path stays within \SI{18.9}{\percent}, edging
a network trained on in-box data (\SI{20.3}{\percent}) while remaining
interpretable and bit-exact (Fig.~\ref{fig:nn}).

\emph{FPGA real-time, HIL, and digital twins.}
FPGAs deliver deterministic real-time in domains adjacent to ours:
IGBT and power-electronics hardware-in-the-loop
\cite{fpga2018igbt,fpga2022hil}, low-latency control
\cite{fpga2020latency,fpga2026optctrl}, and reduced-order digital twins
\cite{fpgadt2024reduction}. These emulate electrical/thermal ODEs or
model-order reductions, not geometric contact mechanics; conversely,
gear-oriented digital twins are dominated by AI fault diagnosis rather than
physics-based contact prediction. To our knowledge no prior FPGA realizes the
geometry-to-pressure gear contact chain.

\emph{Fixed-point geometry queries and open EDA.}
The nearest hardware precedents come from robotics and graphics: fixed-point
ray--triangle intersection \cite{fpga2007raytriangle}, hardware collision and
closest-point queries \cite{collision2004hardware}, and motion-planning
geometry engines \cite{fpga2016motionplan} all demonstrate that pose
transforms, nearest-point search, and gating map cleanly onto fixed-point
logic. We transfer this technique class to gear tooth contact---gap stencil,
baked pseudo-inverse curvature fit, and serialized elliptical Hertz---and,
through a fully open-source flow (Yosys/nextpnr-xilinx/prjxray via openXC7
\cite{shah2019yosys,openxc7}) with no floating-point IP, verify it bit-exact
against golden on retired data-center silicon. This is, to our knowledge, the
first open-source, fixed-point, silicon-verified TCA accelerator.

\section{Background and Problem Formulation}
\label{sec:background}

\emph{Hypoid meshing and ease-off.} Hypoid gears transmit motion between
non-intersecting, non-parallel axes; the hypoid offset introduces lengthwise
sliding and a contact patch that migrates across the flank as the mesh rolls.
The mating flanks are deliberately non-conjugate: a prescribed \emph{ease-off}
---the normal deviation of the pinion/gear surfaces from perfect conjugate
action---localizes contact away from the edges and shapes the transmission
error \cite{litvin2004gear,kolivand2009ease}. Ease-off topography is thus the
design lever for load capacity and NVH, and any digital twin of the mesh must
reproduce its effect at every roll angle.

\emph{From TCA to LTCA.} Unloaded tooth-contact analysis (TCA) locates, at each
mesh phase $\phi$, the instantaneous contact point where the two flanks share a
common normal, tracing the contact path and the unloaded transmission error
$\mathrm{TE}_0(\phi)$. Its primitive is the signed separation of a posed pinion
point $\mathbf{p}=\mathbf{R}\mathbf{p}_0+\mathbf{t}$ from a gear surface point
$\mathbf{g}$ with unit normal $\mathbf{n}$,
\begin{equation}
\gamma \;=\; -\,\mathbf{n}\cdot(\mathbf{p}-\mathbf{g}),
\label{eq:gap}
\end{equation}
minimized over the gear flank (Alg.~\ref{alg:gap}). Under torque the point
contact opens into an elliptical patch shared by several tooth pairs. Loaded TCA
(LTCA) discretizes the patch and couples the nodal load vector $\mathbf{w}$
through half-space influence coefficients $\mathbf{C}$ (Boussinesq kernel
\cite{boussinesq1885,johnson1985contact}); the mesh-phase load distribution and
rigid approach $\delta$ satisfy a linear complementarity problem (LCP)
\begin{equation}
\mathbf{s}=\mathbf{C}\mathbf{w}+\mathbf{h}-\delta\mathbf{1}\ge\mathbf{0},\quad
\mathbf{w}\ge\mathbf{0},\quad \mathbf{w}^{\!\top}\mathbf{s}=0,
\label{eq:lcp}
\end{equation}
where $\mathbf{h}$ is the ease-off/geometric gap vector and $\mathbf{s}$ the
residual separations. Equivalent to a nonnegative least-squares fit,
\eqref{eq:lcp} is solved by a sequential active-set sweep; $\mathbf{C}$ is dense
and near-singular toward the patch boundary, forcing fp64 and dominating the
per-phase cost \cite{icm2018bevel,kolivand2009ease}. Sweeping $\delta(\phi)$
across the mesh yields the loaded transmission error (LTE), the primary
gear-whine excitation \cite{te2018nvh,evgear2025nvh}.

\emph{Problem statement.} A real-time digital twin, or a hardware-in-the-loop
rig, must evaluate contact once per integration step at a \emph{fixed},
jitter-free latency. Full LTCA violates both requirements: the active-set
iteration count is data-dependent, so timing is nondeterministic, and the fp64
dense solve is far too heavy for a per-step budget. Yet cost and conditioning
are strongly heterogeneous across the pipeline. The geometry of \eqref{eq:gap}
---pose transform, gap, nearest-point search---is well-conditioned and tolerates
aggressive quantization: gap error falls below a micrometre at $\ge 20$-bit
coordinates (Fig.~\ref{fig:precision}), so it needs no fp64. Local curvature and
the elliptical-Hertz peak pressure then follow from cheap closed forms. Only the
coupled contact law \eqref{eq:lcp} genuinely demands fp64 and iteration. This
asymmetry frames the central question of the paper: \emph{which stages of the
contact computation admit deterministic, fixed-point hardware evaluation, and
where must the LCP be retained?} We answer it with a physics-direct CPU/FPGA
partition (Fig.~\ref{fig:partition}) executed under a fully fixed-point data
contract (Table~\ref{tab:fixedpoint}). The hardware returns a \emph{preview}
peak-pressure and TE estimate at bounded, deterministic latency---an
engineering-magnitude surrogate, not the full LCP---while the fp64 solve is
reserved for the edge-loaded pressure peaks that only it resolves.

\begin{table}[!t]
\caption{Fully Fixed-Point Data Contract (No Floating-Point IP)}
\label{tab:fixedpoint}
\centering
\footnotesize
\begin{tabular}{lll}
\toprule
Signal & Format & Meaning \\
\midrule
coordinate      & Q(24,16) & position (\si{\milli\meter}) \\
normal          & Q1.19    & unit normal component \\
load $W$        & Q(32,8)  & contact load (\si{\newton}) \\
curvature $A_{-}, B_{-}$ & Q(32,20) & principal curvatures (\si{\per\milli\meter}) \\
semi-axes $a, b$ & Q(32,20) & contact ellipse semi-axes (\si{\milli\meter}) \\
$p_{\max}$      & Q(32,8)  & peak Hertzian pressure (\si{\mega\pascal}) \\
POW/CBRT LUT    & 256-entry & power-of-2 step (shift interp.) \\
\bottomrule
\end{tabular}
\end{table}

\section{CPU/FPGA Partitioning}
\label{sec:partition}

We split the tooth-contact computation by two orthogonal hardware-admissibility
tests, not by which stage looks expensive (Fig.~\ref{fig:partition}). A stage
earns a place in the fabric only if it is (i)~\emph{throughput-dominated under a
static schedule}---a fixed cycle count with no data-dependent control flow---and
(ii)~\emph{well conditioned in bounded fixed-point}, since the open-source flow
(openXC7~\cite{openxc7,shah2019yosys}) exposes no hardware floating-point IP.
Work that fails either test---sequential, branch-divergent, or
ill-conditioned---stays on the host CPU. This criterion, not raw operation count,
fixes the split.

The physics-direct pipeline passes both tests at every stage. The Stage~A/B
geometry kernel is a nearest-point query: transform the pinion point by the rigid
pose, project the pinion--gear difference onto the gear normal for a signed gap,
and $\perp^{2}$-gate an arg-min over the candidate cloud
(Alg.~\ref{alg:gap}). This is structurally the ray/closest-primitive query that
reconfigurable hardware has long streamed in fixed-point at one candidate per lane
per cycle~\cite{fpga2007raytriangle,collision2004hardware,fpga2016motionplan}:
$L$ lanes, initiation interval~$1$, a $\log_{2}L$ reduction tree, and no branch
that depends on the data. Stage~C collapses the local ease-off least-squares fit
to a \emph{baked pseudo-inverse}: with a fixed stencil the normal-equation
operator $M=(A^{\top}A)^{-1}A^{\top}$ is a compile-time constant, so curvature
fitting becomes a fixed-weight multiply--accumulate---a two-dimensional
Savitzky--Golay convolution~\cite{savitzky1964smoothing}---followed by a
$2{\times}2$ closed-form eigen-solve for the relative principal curvatures
(Alg.~\ref{alg:fit}). Stage~D is the closed-form Hamrock--Brewe elliptical Hertz
solution~\cite{hamrock1977elliptical}, serialized onto a single iterative divider
(Alg.~\ref{alg:bhseq}). All three carry static schedules and map to deterministic
$\sigma=0$ datapaths (Table~\ref{tab:latency}).

The exact loaded solve resists both tests. Full loaded tooth contact analysis
(LTCA) reduces to a linear complementarity problem (LCP) solved by an NNLS
active-set iteration whose control
flow is intrinsically sequential: each pivot adds or drops a contact index from
the sign pattern of the previous partial solution, so the branch taken at step
$k$ is unknown until step $k{-}1$ resolves. In our profiling this data-dependent
recursion dominates the solve, exceeding \SI{95}{\percent} of its run time, so by
Amdahl's law hardening the parallel remainder buys at most $\approx 1/0.95 \approx
1.05\times$---the fabric would gain essentially nothing.
Independently, the Boussinesq influence matrix~\cite{boussinesq1885} is dense and
near-singular---its $1/r$ kernel worsens in conditioning as the mesh refines---and
the active-set pivots stay stable only in fp64, precisely the arithmetic the
fabric lacks. The exact solve therefore remains on the CPU
(\SI{2005}{\milli\second}, \SI{2}{\percent} jitter), run offline to \emph{anchor}
the on-chip loaded preview rather than to feed it per step.

Even the fabric-friendly geometry carries a ceiling that dictates the
architecture. A brute-force contact search is $O(N^{2})$ per mesh phase---each of
$N$ pinion points scanned against $N$ gear candidates---while the saturated
accelerator sustains at most $L\,f_{\mathrm{clk}}$ candidate gaps per second, i.e.
$63\times\SI{50}{\mega\hertz}\approx\num{3.15e9}$ evaluations/s at the pipelined
target (the $L\!\approx\!63$ DSP wall of Fig.~\ref{fig:throughput}), four-fold
lower on the \SI{12.5}{\mega\hertz} board. Resolving loaded transmission error
(LTE) across the NVH band demands a mesh-phase update rate $f_{\phi}$ for which
$N^{2}f_{\phi}$ overruns this budget at any realistic cloud size: re-deriving the
full geometry every real-time step does not fit---on this chip or a larger one.
The resolution is architectural. Under fixed kinematics the contact path, gap, and
relative curvatures are smooth functions of a single scalar---the roll/mesh
phase---so they are precomputed once on the host, tabulated, and downloaded as a
few-hundred-byte phase table (Fig.~\ref{fig:partition}); the gap kernel is
retained only as the primitive for bounded, local refinement, never a full-cloud
re-search per step. The FPGA then spends its operation budget on the
load-dependent Hertz preview, which actually changes with torque, and
\emph{evaluates} the invariant geometry instead of rediscovering it. The rule is
blunt: do not recompute geometry every step; evaluate a well-budgeted phase
function.

\section{Fixed-Point Contact Pipeline}
\label{sec:pipeline}

The engine evaluates one pinion query point against the mating gear cloud in four
feed-forward stages (overview in Fig.~\ref{fig:partition}; microarchitectures in
Figs.~\ref{fig:gapcore} and~\ref{fig:bhseq}): rigid pose transform and signed gap
(A/B), a baked quadratic curvature fit (C), and an elliptic-Hertz load response (D).
The governing principle is \emph{physics-direct} arithmetic in which every runtime
unknown is resolved to a fixed, data-independent operation count: no matrix is
inverted on device, no floating-point IP is instantiated, and no branch depends on
operand values. Latency is therefore a constant rather than a distribution
(Table~\ref{tab:latency}), and the datapath obeys one fixed-point contract
(Table~\ref{tab:fixedpoint})---coordinates in Q(24,16)\,\si{\milli\meter}, unit
normals in Q1.19, curvatures in Q(32,20), pressure in Q(32,8)---with no fp64 anywhere.

\subsection{Stages A/B: Geometric Contact Gap}
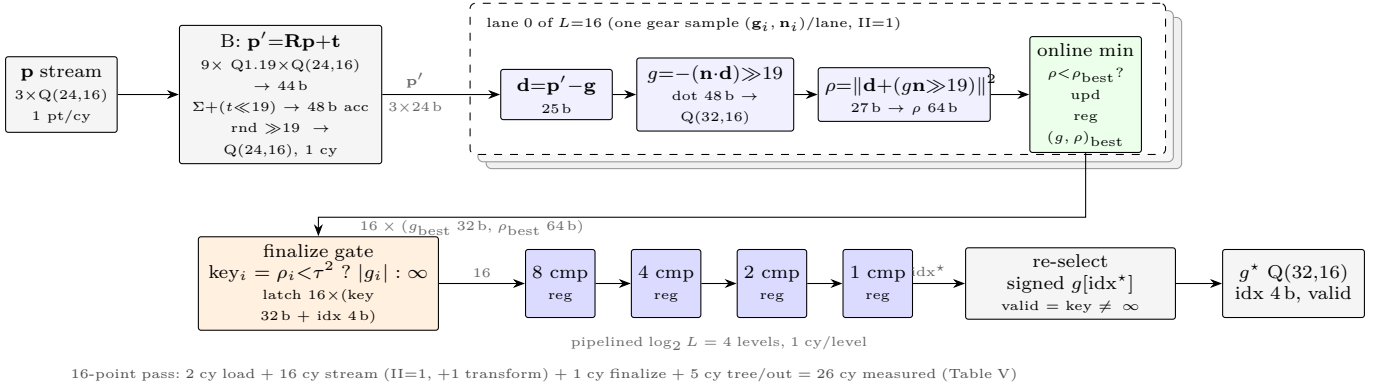
\begin{figure*}[!t]
  \centering
  \begin{tikzpicture}[
    font=\scriptsize,
    >={Stealth[length=1.6mm]},
    blk/.style={draw, rounded corners=1pt, align=center, inner sep=2.5pt, fill=black!4},
    lane/.style={blk, fill=blue!6},
    buslbl/.style={font=\tiny, text=black!60, inner sep=1pt},
  ]
    \node[blk, minimum height=10mm, text width=13mm] (inp) at (0,0)
      {$\mathbf{p}$ stream\\{\tiny $3{\times}$Q(24,16)}\\{\tiny 1 pt/cy}};
    \node[blk, minimum height=10mm, text width=25mm] (xf) at (2.9,0)
      {\textbf{B: $\mathbf{p}'{=}\mathbf{Rp}{+}\mathbf{t}$}\\
       {\tiny $9{\times}$ Q1.19${\times}$Q(24,16) $\to$ 44\,b}\\
       {\tiny $\Sigma{+}(t{\ll}19) \to$ 48\,b acc}\\
       {\tiny rnd ${\gg}19 \to$ Q(24,16), 1 cy}};
    \draw[rounded corners=2pt, fill=black!6, draw=black!40] (5.62,-0.98) rectangle (14.82,1.02);
    \draw[rounded corners=2pt, fill=black!3, draw=black!40] (5.51,-0.89) rectangle (14.71,1.11);
    \node[draw, dashed, rounded corners=2pt, fill=white, inner sep=0pt,
          minimum width=92mm, minimum height=20mm, anchor=south west] (lanebox) at (5.4,-0.8) {};
    \node[anchor=north west, font=\tiny\itshape] at (5.5,1.15) {lane 0 of $L{=}16$ (one gear sample $(\mathbf{g}_i,\mathbf{n}_i)$/lane, II$=$1)};
    \node[lane, text width=13mm] (d) at (6.55,0)
      {$\mathbf{d}{=}\mathbf{p}'{-}\mathbf{g}$\\{\tiny 25\,b}};
    \node[lane, text width=19mm] (sd) at (8.65,0)
      {$g{=}{-}(\mathbf{n}{\cdot}\mathbf{d}){\gg}19$\\{\tiny dot 48\,b $\to$ Q(32,16)}};
    \node[lane, text width=21mm] (pp) at (11.15,0)
      {$\rho{=}\lVert\mathbf{d}{+}(g\mathbf{n}{\gg}19)\rVert^{2}$\\{\tiny 27\,b $\to$ $\rho$ 64\,b}};
    \node[lane, fill=green!8, text width=13mm] (mn) at (13.55,0)
      {online min\\{\tiny $\rho{<}\rho_{\mathrm{best}}$? upd}\\{\tiny reg $(g,\rho)_{\mathrm{best}}$}};
    \draw[->] (inp) -- (xf);
    \draw[->] (xf) -- node[buslbl, above=0.5mm, pos=0.28]{$\mathbf{p}'$} node[buslbl, below=0.5mm, pos=0.28]{$3{\times}$24\,b} (d);
    \draw[->] (d) -- (sd);
    \draw[->] (sd) -- (pp);
    \draw[->] (pp) -- (mn);
    \node[blk, fill=orange!12, text width=30mm, minimum height=9mm] (gate) at (3.4,-2.5)
      {\textbf{finalize gate}\\
       key$_i$ = $\rho_i{<}\tau^2$ ? $|g_i|$ : $\infty$\\
       {\tiny latch $16{\times}$(key 32\,b $+$ idx 4\,b)}};
    \node[blk, fill=blue!14, minimum height=9mm] (c1) at (6.6,-2.5) {8 cmp\\{\tiny reg}};
    \node[blk, fill=blue!14, minimum height=9mm] (c2) at (8.0,-2.5) {4 cmp\\{\tiny reg}};
    \node[blk, fill=blue!14, minimum height=9mm] (c3) at (9.4,-2.5) {2 cmp\\{\tiny reg}};
    \node[blk, fill=blue!14, minimum height=9mm] (c4) at (10.8,-2.5) {1 cmp\\{\tiny reg}};
    \node[blk, text width=26mm, minimum height=9mm] (rsel) at (13.35,-2.5)
      {re-select signed $g[\mathrm{idx}^\star]$\\{\tiny valid $=$ key $\ne\infty$}};
    \node[blk, text width=17mm, minimum height=9mm] (outp) at (16.3,-2.5)
      {$g^\star$ Q(32,16)\\idx 4\,b, valid};
    \draw[->] (mn.south) -- ++(0,-0.85) -| node[buslbl, pos=0.4, below]
      {$16\times(g_{\mathrm{best}}$ 32\,b, $\rho_{\mathrm{best}}$ 64\,b$)$} (gate.north);
    \draw[->] (gate) -- node[buslbl, above=0.5mm]{$16$} (c1);
    \draw[->] (c1) -- (c2);
    \draw[->] (c2) -- (c3);
    \draw[->] (c3) -- (c4);
    \draw[->] (c4) -- node[buslbl, above=0.5mm, pos=0.3]{idx$^\star$} (rsel);
    \draw[->] (rsel) -- (outp);
    \node[buslbl, anchor=north] at (8.7,-3.15) {pipelined $\log_2 L=4$ levels, 1 cy/level};
    \node[anchor=west, font=\tiny, text=black!70] at (0,-3.7)
      {16-point pass: 2 cy load $+$ 16 cy stream (II$=$1, $+1$ transform) $+$ 1 cy finalize
       $+$ 5 cy tree/out $=$ \textbf{26 cy measured} (Table~\ref{tab:latency})};
  \end{tikzpicture}
  \caption{Stage A/B gap-kernel microarchitecture with fixed-point word growth.
  One shared transform unit (B) streams pinion points at one per cycle
  ($9$ multiplies Q1.19${\times}$Q(24,16)${\to}$\SI{44}{b}, \SI{48}{b}
  accumulate, round-shift back to Q(24,16)); the point is broadcast to
  $L{=}16$ lanes, each holding one gear sample and updating an online
  minimum of the perpendicular residual $\rho$ (\SI{64}{b}) at initiation
  interval~1---no candidate list is ever stored. On \texttt{finalize}, a
  perpendicular gate replaces each lane key by $|g|$ or $\infty$
  (Eq.~\ref{eq:argmin}), and a registered $\log_2 L$-level comparator tree
  reduces the $16$ keys in $4$ cycles; the winning index re-selects the
  signed gap. All rounding is round-half-up shifting; no intermediate
  exceeds \SI{64}{b} and no operation count depends on data.}
  \label{fig:gapcore}
\end{figure*}

A rigid pose maps the pinion point into the gear frame, $\mathbf{p}'=\mathbf{R}\mathbf{p}+\mathbf{t}$.
For each gear sample $(\mathbf{g}_i,\mathbf{n}_i)$ the signed gap along the surface
normal and the off-normal (tangential) residual are
\begin{equation}
  g_i=-\,\mathbf{n}_i\!\cdot\!(\mathbf{p}'-\mathbf{g}_i),\qquad
  \rho_i=\big\lVert(\mathbf{p}'-\mathbf{g}_i)+g_i\,\mathbf{n}_i\big\rVert^{2}.
  \label{eq:gaplane}
\end{equation}
The scalar $g_i$ is the quantity of interest, but selecting the nearest sample by
$|g_i|$ alone is unreliable near tooth edges: a gear point lying laterally off the
query still projects to a small $|g_i|$ while sitting far from the normal line.
The term $\rho_i=\lVert\mathbf{d}-(\mathbf{n}_i\!\cdot\!\mathbf{d})\mathbf{n}_i\rVert^{2}$,
$\mathbf{d}=\mathbf{p}'-\mathbf{g}_i$, is exactly that lateral offset squared, and a
gate $\rho_i<\tau^{2}$ discards such spurious tangential candidates before the
selection. The mating point is the gated arg-min
\begin{equation}
  i^{\star}=\operatorname*{arg\,min}_{\,i:\ \rho_i<\tau^{2}}\ |g_i|,\qquad
  g^{\star}=g_{i^{\star}}.
  \label{eq:argmin}
\end{equation}
The cloud is striped round-robin over $L$ lanes, each consuming one sample per cycle
at II$=1$; every lane retains its own $\rho$-minimizer, and a $\log_2 L$-level tree
carries out the gated cross-lane reduction (Alg.~\ref{alg:gap}; microarchitecture
and word growth in Fig.~\ref{fig:gapcore}). Lacking any
data-dependent control, the kernel retires a query every cycle at a fixed 26-cycle
latency.

\subsection{Stage C: Baked Curvature Fit}
Around $i^{\star}$ a fixed stencil $\{(s_i,t_i)\}$ in the local tangent frame samples
the gap field, so the quadratic design matrix $\mathbf{A}$ (columns
$1,\,s,\,t,\,s^{2},\,st,\,t^{2}$) is a compile-time constant. Its Moore--Penrose
pseudo-inverse is formed once on the host and quantized into ROM, reducing the fit to
a fixed-weight multiply--accumulate,
\begin{equation}
  \mathbf{M}=(\mathbf{A}^{\!\top}\mathbf{A})^{-1}\mathbf{A}^{\!\top},\qquad
  \mathbf{c}=\mathbf{M}\,\mathbf{g}.
  \label{eq:mac}
\end{equation}
The six coefficients of the local gap surface
$\delta(s,t)=c_0+c_1 s+c_2 t+c_3 s^{2}+c_4 st+c_5 t^{2}$ are six inner products
against constant rows of $\mathbf{M}$; no matrix inversion or division runs at
runtime (Alg.~\ref{alg:fit}). This is precisely a two-dimensional Savitzky--Golay
estimator---a precomputed least-squares convolution over a fixed
stencil~\cite{savitzky1964smoothing}---specialized to a curved tooth surface. The
relative principal curvatures are the eigenvalues of the fitted quadratic form; for
the $2\times2$ symmetric Hessian $\mathbf{H}=\big[\begin{smallmatrix}2c_3 & c_4\\
c_4 & 2c_5\end{smallmatrix}\big]$ they are available in closed form,
\begin{equation}
  \{A_-,\,B_-\}=\tfrac12\Big[(c_3+c_5)\pm\sqrt{(c_3-c_5)^{2}+c_4^{2}}\Big],
  \label{eq:eig}
\end{equation}
so a single non-restoring integer square root (shift-and-add, no divide) replaces an
iterative eigensolver.

\subsection{Stage D: Load-Dependent Elliptic Hertz}
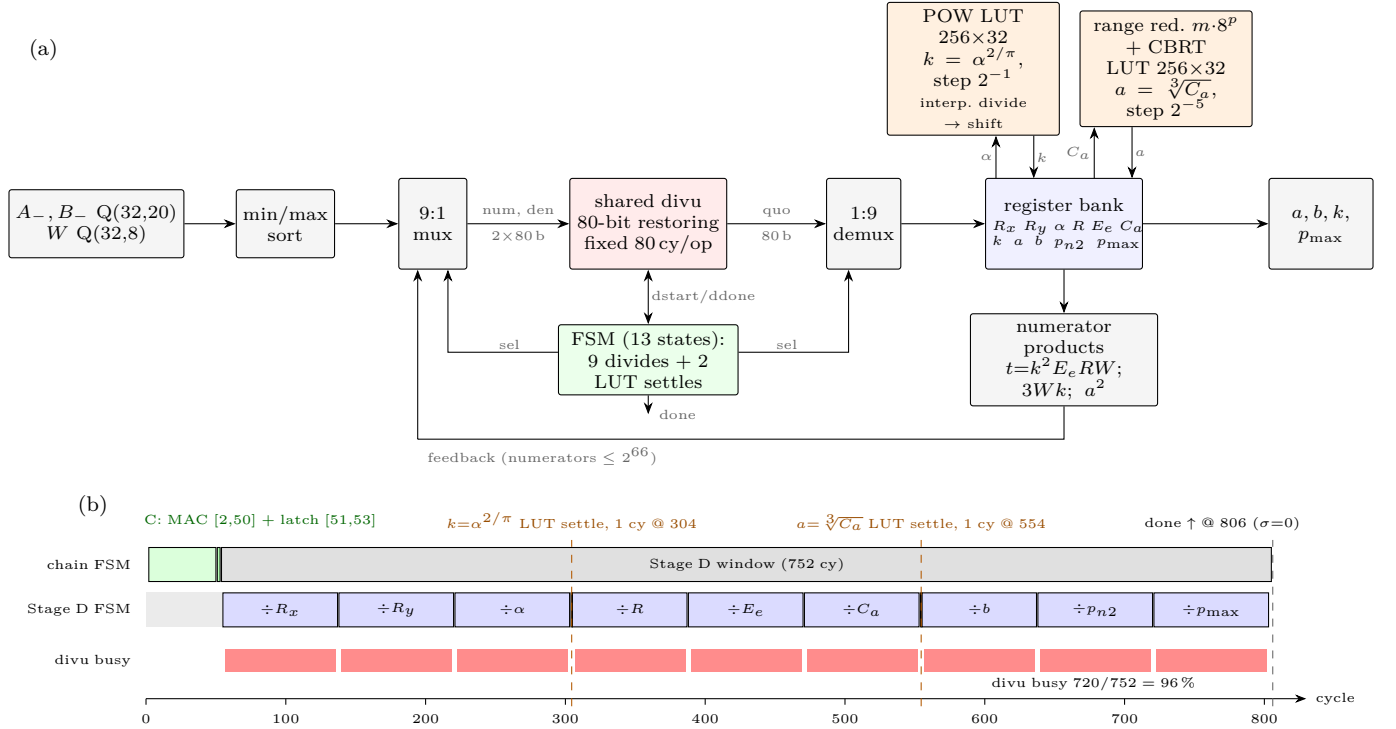
\begin{figure*}[!t]
  \centering
  \begin{tikzpicture}[
    font=\scriptsize,
    >={Stealth[length=1.6mm]},
    blk/.style={draw, rounded corners=1pt, align=center, inner sep=2.5pt, fill=black!4},
    buslbl/.style={font=\tiny, text=black!60, inner sep=1pt},
  ]
    \node[blk, minimum height=9mm] (inp) at (0.0,0)  {$A_-,B_-$ Q(32,20)\\$W$ Q(32,8)};
    \node[blk, minimum height=9mm] (mm)  at (2.5,0) {min/max\\sort};
    \node[blk, minimum height=12mm, minimum width=9mm] (mux) at (4.45,0) {9:1\\mux};
    \node[blk, minimum height=12mm, fill=red!8, text width=19mm] (div) at (7.3,0)
      {\textbf{shared \texttt{divu}}\\80-bit restoring\\fixed 80\,cy/op};
    \node[blk, minimum height=12mm, minimum width=9mm] (dmx) at (10.15,0) {1:9\\demux};
    \node[blk, minimum height=12mm, fill=blue!6, text width=19mm] (rb) at (12.8,0)
      {register bank\\[-1pt]{\tiny $R_x\;R_y\;\alpha\;R\;E_e\;C_a$}\\[-2pt]
       {\tiny $k\;\;a\;\;b\;\;p_{n2}\;\;p_{\max}$}};
    \node[blk, minimum height=12mm, text width=12mm] (out) at (16.2,0)
      {$a,b,k,$\\$p_{\max}$};
    \draw[->] (inp) -- (mm);
    \draw[->] (mm) -- (mux);
    \draw[->] (mux) -- node[buslbl, above=0.5mm]{num, den} node[buslbl, below=0.5mm]{$2{\times}$80\,b} (div);
    \draw[->] (div) -- node[buslbl, above=0.5mm]{quo} node[buslbl, below=0.5mm]{80\,b} (dmx);
    \draw[->] (dmx) -- (rb);
    \draw[->] (rb) -- (out);
    \node[blk, fill=orange!12, text width=21mm] (pow) at (11.6,2.05)
      {POW LUT $256{\times}32$\\$k=\alpha^{2/\pi}$, step $2^{-1}$\\
       {\tiny interp.\ divide $\to$ shift}};
    \node[blk, fill=orange!12, text width=21mm] (cbrt) at (14.15,2.05)
      {range red.\ $m{\cdot}8^{p}$\\$+$ CBRT LUT $256{\times}32$\\
       $a=\sqrt[3]{C_a}$, step $2^{-5}$};
    \draw[->] (11.9,0.6) -- node[buslbl, left]{$\alpha$}  (11.9,0 |- pow.south);
    \draw[<-] (12.4,0.6) -- node[buslbl, right]{$k$}   (12.4,0 |- pow.south);
    \draw[->] (13.2,0.6) -- node[buslbl, left]{$C_a$}  (13.2,0 |- cbrt.south);
    \draw[<-] (13.7,0.6) -- node[buslbl, right]{$a$}   (13.7,0 |- cbrt.south);
    \node[blk, fill=green!8, text width=22mm, minimum height=9mm] (fsm) at (7.3,-1.8)
      {\textbf{FSM} (13 states):\\9 divides $+$ 2 LUT settles};
    \draw[<->] (fsm) -- node[buslbl, right]{\texttt{dstart}/\texttt{ddone}} (div);
    \draw[->] ($(fsm.west)+(0,1mm)$) -| node[buslbl, pos=0.22, above]{sel}
      ($(mux.south)+(2mm,0)$);
    \draw[->] ($(fsm.east)+(0,1mm)$) -| node[buslbl, pos=0.22, above]{sel}
      ($(dmx.south)+(-2mm,0)$);
    \draw[->] (fsm.south) -- ++(0,-0.25) node[buslbl, right=1mm]{\texttt{done}};
    \node[blk, text width=23mm, minimum height=9mm] (prod) at (12.8,-1.8)
      {numerator products\\$t{=}k^{2}E_eRW$;\; $3Wk$;\; $a^{2}$};
    \draw[->] (rb.south) -- (prod.north);
    \draw[->] (prod.south) |- (4.25,-2.85) (4.25,-2.85) -- ($(mux.south)+(-2mm,0)$);
    \node[buslbl, anchor=north] at (5.9,-2.92) {feedback (numerators $\le 2^{66}$)};
    \node[font=\bfseries\footnotesize] at (-0.7,2.3) {(a)};
  \end{tikzpicture}

  \vspace{2.5mm}

  \begin{tikzpicture}[font=\scriptsize, x=0.0185cm, >={Stealth[length=1.5mm]}]
    \draw[->] (0,0) -- (830,0) node[right, font=\tiny]{cycle};
    \foreach \x in {0,100,...,800}{
      \draw (\x,0) -- (\x,-0.08) node[below, font=\tiny]{\x};}
    \def\rowc{0.35}
    \foreach \s/\e in {57/136,140/219,223/302,307/386,390/469,473/552,557/636,640/719,723/802}{
      \fill[red!45] (\s,\rowc) rectangle (\e,\rowc+0.3);}
    \node[anchor=east, font=\tiny] at (-4,\rowc+0.15) {\texttt{divu} busy};
    \def\rowb{0.95}
    \fill[black!8] (0,\rowb) rectangle (54,\rowb+0.45);
    \foreach \s/\e/\l in {55/137/{\div R_x},138/220/{\div R_y},221/303/{\div\alpha},
                          305/387/{\div R},388/470/{\div E_e},471/553/{\div C_a},
                          555/637/{\div b},638/720/{\div p_{n2}},721/803/{\div p_{\max}}}{
      \draw[fill=blue!14] (\s,\rowb) rectangle (\e,\rowb+0.45);
      \node at ({(\s+\e)/2},\rowb+0.225) {\tiny$\l$};}
    \draw[fill=orange!80] (304,\rowb) rectangle (305,\rowb+0.45);
    \draw[fill=orange!80] (554,\rowb) rectangle (555,\rowb+0.45);
    \node[anchor=east, font=\tiny] at (-4,\rowb+0.225) {Stage D FSM};
    \def\rowa{1.55}
    \draw[fill=green!14] (2,\rowa) rectangle (50,\rowa+0.45);
    \draw[fill=green!30] (51,\rowa) rectangle (53,\rowa+0.45);
    \draw[fill=black!12] (54,\rowa) rectangle (805,\rowa+0.45);
    \node at (430,\rowa+0.225) {\tiny Stage D window (752 cy)};
    \node[anchor=east, font=\tiny] at (-4,\rowa+0.225) {chain FSM};
    \draw[dashed, orange!70!black] (304.5,-0.1) -- (304.5,\rowa+0.55);
    \node[above, font=\tiny, align=center, text=orange!60!black] at (304.5,\rowa+0.55)
      {$k{=}\alpha^{2/\pi}$ LUT settle, 1 cy @ 304};
    \draw[dashed, orange!70!black] (554.5,-0.1) -- (554.5,\rowa+0.55);
    \node[above, font=\tiny, align=center, text=orange!60!black] at (554.5,\rowa+0.55)
      {$a{=}\sqrt[3]{C_a}$ LUT settle, 1 cy @ 554};
    \draw[dashed, black!60] (806,-0.1) -- (806,\rowa+0.55);
    \node[above, font=\tiny] at (770,\rowa+0.55) {\texttt{done} $\uparrow$ @ 806 ($\sigma{=}0$)};
    \node[anchor=west, font=\tiny, text=green!40!black] at (-8,\rowa+0.80)
      {C: MAC [2,50] $+$ latch [51,53]};
    \node[anchor=west, font=\tiny, align=left] at (598,0.2)
      {\texttt{divu} busy $720/752$ = \SI{96}{\percent}};
    \node[font=\bfseries\footnotesize] at (-38,2.55) {(b)};
  \end{tikzpicture}
  \caption{Stage D serialized Hamrock--Brewe core. (a)~Microarchitecture: a
  single 80-bit restoring divider (\texttt{divu}, fixed \num{80} cycles per
  operation) is time-multiplexed by a 13-state FSM across all nine divisions
  ($R_x, R_y, \alpha, R, E_e, C_a, b, p_{n2}, p_{\max}$); the two
  transcendentals resolve in one cycle each through 256-entry LUTs whose
  power-of-two step sizes ($2^{-1}$, $2^{-5}$) turn interpolation division
  into a shift, and the cube root is range-reduced to $m\in[1,8)$ before
  lookup. Numerator products are formed incrementally so no intermediate
  exceeds $2^{66}$. (b)~Cycle-accurate schedule measured from RTL simulation
  of the full CSR chain: after a 49-cycle Stage C MAC stream and a 3-cycle
  curvature latch, each division occupies exactly $80{+}3$ handshake cycles
  (blue); the two LUT settles take one cycle each (orange); divider occupancy
  is $720/752$ cycles (\SI{96}{\percent}) and \texttt{done} rises at cycle
  \num{806} with zero jitter.}
  \label{fig:bhseq}
\end{figure*}

Given the load $W$ and the curvatures ordered $A_-\!\le\!B_-$, Stage D returns the
Hamrock--Brewe elliptic-Hertz solution~\cite{hamrock1977elliptical,johnson1985contact}.
With reduced radii $R_x=1/(2A_-)$, $R_y=1/(2B_-)$, effective radius
$1/R=1/R_x+1/R_y$, curvature ratio $\alpha=R_x/R_y$, ellipticity $k=\alpha^{2/\pi}$,
and second-kind elliptic-integral approximation $\mathcal{E}\!\approx\!1+(\pi/2-1)/\alpha$,
\begin{equation}
  a=\Big(\tfrac{6\,k^{2}\mathcal{E}\,W R}{\pi E'}\Big)^{1/3},\qquad
  b=\frac{a}{k},\qquad
  p_{\max}=\frac{3W}{2\pi a b},
  \label{eq:hertz}
\end{equation}
where $E'=2E^{\ast}$ is the effective modulus. Cast this way the entire ellipse costs
exactly one cube root: $b$ follows from $a$ by the closed ratio $b=a/k$, and
$p_{\max}$ from $a,b$. The serialized datapath (Alg.~\ref{alg:bhseq},
microarchitecture and measured schedule in Fig.~\ref{fig:bhseq})
time-multiplexes the nine divisions onto a single 80-bit iterative divider, reduces
the cube-root argument to a fixed interval before a table lookup, and lays the
transcendental tables ($(\cdot)^{2/\pi}$, $\sqrt[3]{\cdot}$) on power-of-two strides
so their interpolation is a shift, not a divide. A representative probed point
resolves end-to-end to $a=0.909$\,\si{\milli\meter}, $b=0.452$\,\si{\milli\meter},
$k=2.01$, and $p_{\max}=1162.5$\,\si{\mega\pascal}. This peak pressure is an
elliptic-Hertz \emph{preview}---an engineering-magnitude estimate at the mating
point, not a full LCP load distribution over the contact patch; its deviation from a
Boussinesq$+$LCP reference is quantified in Fig.~\ref{fig:accuracy}. Because the
operation count is fixed, the serialized Hertz completes in 756 cycles and the full
C$\rightarrow$D chain in 806 cycles with zero jitter ($\sigma=0$), while a fully
combinational preview trims the chain to 58 cycles (Table~\ref{tab:latency}).

\begin{algorithm}[!t]
  \caption{Geometric contact-gap kernel: $\perp^{2}$-gated nearest-point search
    over $L$ lanes with cross-lane arg-min by signed gap. Single initiation
    interval (II${=}1$), fixed-point datapath only (no fp64).}
  \label{alg:gap}
  \begin{algorithmic}[1]
    \Require gear cloud $\{(\mathbf{g}_i,\mathbf{n}_i)\}_{i=0}^{N-1}$, laid out
             round-robin over $L$ lanes ($i=l+kL$)
    \Require pinion point $\mathbf{p}$, rigid pose $(\mathbf{R},\mathbf{t})$,
             gate radius $\tau^{2}$
    \Ensure  signed gap $g^{\star}$ and gear index $i^{\star}$ of the mating point
    \Statex \textit{All arithmetic in fixed-point Q-format; one gear point per
      lane per cycle (II${=}1$).}
    \State $\mathbf{p}' \gets \mathbf{R}\,\mathbf{p} + \mathbf{t}$
           \Comment{pose transform (Stage A)}
    \ForAll{lanes $l \in \{0,\dots,L-1\}$ \textbf{in parallel}}
      \State $(\rho_l,\, s_l,\, i_l) \gets (+\infty,\, +\infty,\, -1)$
             \Comment{per-lane running best}
    \EndFor
    \For{$k \gets 0$ \textbf{to} $\lceil N/L\rceil-1$}
        \Comment{stream gear points, pipelined II${=}1$}
      \ForAll{lanes $l$ \textbf{in parallel}}
          \Comment{broadcast $\mathbf{p}'$}
        \State $i \gets l + kL$
        \State $\mathbf{d} \gets \mathbf{p}' - \mathbf{g}_i$
        \State $s \gets -\,\mathbf{n}_i \cdot \mathbf{d}$
               \Comment{signed gap along normal}
        \State $\rho \gets \lVert \mathbf{d} + s\,\mathbf{n}_i \rVert^{2}$
               \Comment{off-normal $\perp^{2}$ residual}
        \If{$\rho < \rho_l$}
            \Comment{running-min $\operatorname*{arg\,min}_{\perp}$}
          \State $(\rho_l,\, s_l,\, i_l) \gets (\rho,\, s,\, i)$
        \EndIf
      \EndFor
    \EndFor
    \ForAll{lanes $l$ \textbf{in parallel}}
      \State $v_l \gets [\,\rho_l < \tau^{2}\,]$
             \Comment{gate: retain near-surface lanes}
    \EndFor
    \State $l^{\star} \gets \displaystyle\operatorname*{arg\,min}_{\,l\,:\,v_l}\,|s_l|$
           \Comment{$\log_{2}L$-level tree reduction}
    \State $(g^{\star},\, i^{\star}) \gets (s_{l^{\star}},\, i_{l^{\star}})$
           \Comment{sentinel if no lane passes the gate}
    \State \Return $(g^{\star},\, i^{\star})$
           \Comment{\texttt{min\_gap}, \texttt{min\_idx}}
  \end{algorithmic}
\end{algorithm}

\begin{algorithm}[!t]
\caption{Stage~C curvature fit: the least-squares design matrix is inverted
  once offline and baked as a constant pseudo-inverse, so on device the
  quadratic fit reduces to a fixed-weight multiply--accumulate followed by a
  closed-form $2{\times}2$ eigen-solve---no matrix inversion or division runs
  at runtime.}
\label{alg:fit}
\begin{algorithmic}[1]
\Statex \textbf{Offline (host, one-time per stencil)}
\Require Fixed stencil $\{(s_i,t_i)\}_{i=0}^{N-1}$ in the local tangent frame
\State $\mathbf{A}\gets[\,\mathbf{1},\ \mathbf{s},\ \mathbf{t},\ \mathbf{s}^{2},\ \mathbf{s}\!\odot\!\mathbf{t},\ \mathbf{t}^{2}\,]\in\mathbb{R}^{N\times6}$
       \Comment{row $i=[\,1,\,s_i,\,t_i,\,s_i^{2},\,s_i t_i,\,t_i^{2}\,]$}
\State $\mathbf{M}\gets(\mathbf{A}^{\!\top}\mathbf{A})^{-1}\mathbf{A}^{\!\top}\in\mathbb{R}^{6\times N}$
       \Comment{Moore--Penrose pseudo-inverse, double precision}
\State quantize $\mathbf{M}$ to Q$(32,20)$ and bake into constant ROM
       \Comment{sole matrix inversion---never repeated on device}
\Statex
\Statex \textbf{Online (FPGA, per contact point)}
\Require gap vector $\mathbf{g}=[\,g_0,\dots,g_{N-1}\,]^{\!\top}$, Q$(24,16)$; baked $\mathbf{M}$
\Ensure relative principal curvatures $A_{-},B_{-}$, Q$(32,20)$
\For{$j\gets0$ \textbf{to} $5$}
       \Comment{6 sequential MACs, $N$ taps each; no runtime inverse}
  \State $c_j\gets\displaystyle\sum_{i=0}^{N-1} M_{ji}\,g_i$
\EndFor
\State $(a,b,c)\gets(c_3,c_4,c_5)$
       \Comment{quadratic coefficients of $s^{2},\,st,\,t^{2}$}
\Statex \hfill$\displaystyle\mathbf{H}=\begin{bmatrix}2a & b\\ b & 2c\end{bmatrix}$
       \quad(Hessian of the fitted gap)
\State $h\gets a+c$
       \Comment{half-trace $=\tfrac12\operatorname{tr}\mathbf{H}$, Q$(32,20)$}
\State $d\gets(a-c)^{2}+b^{2}$
       \Comment{$\ge0$; squares double the fraction $\Rightarrow$ Q$(64,40)$}
\State $r\gets\Call{Isqrt}{d}$
       \Comment{non-restoring integer $\sqrt{\cdot}$; halves fraction back to Q$(32,20)$}
\State $A_{-}\gets(h+r)\gg1$;\quad $B_{-}\gets(h-r)\gg1$
       \Comment{eigenvalues $\lambda_{\pm}=h\pm r$; curvature $=\lambda/2$ via arithmetic shift}
\State \Return $A_{-},\,B_{-}$
\Statex
\Function{Isqrt}{$d$} \Comment{non-restoring binary square root: shift and add/subtract only}
  \State $r\gets0$;\quad $b\gets 1\ll62$ \Comment{$b$: top power of $4$ in the $64$-bit datum}
  \While{$b>d$}
    \State $b\gets b\gg2$
  \EndWhile
  \While{$b\neq0$}
    \If{$d\ge r+b$}
      \State $d\gets d-(r+b)$;\quad $r\gets(r\gg1)+b$
    \Else
      \State $r\gets r\gg1$
    \EndIf
    \State $b\gets b\gg2$
  \EndWhile
  \State \Return $r$ \Comment{$=\lfloor\sqrt{d}\rfloor$; no multiply, no divide}
\EndFunction
\end{algorithmic}
\end{algorithm}

\begin{algorithm}[!t]
\caption{Serialized Hamrock--Brewe elliptic Hertz preview (Stage~D): one
time-shared 80-bit iterative divider, two power-of-two-stride lookup tables,
and a single cube root. Deterministic \num{756} cycles (Stage~D); \num{806}
cycles for the full C$\rightarrow$D chain, zero jitter.}
\label{alg:bhseq}
\begin{algorithmic}[1]
\Require Relative principal curvatures $A_{\min}\!\le\!B_{\max}$, format Q(32,20)
  [\si{\per\milli\meter}]; contact load $W$, format Q(32,8) [\si{\newton}];
  effective modulus $E'\!=\!2E^{\ast}$; tables $\mathrm{POW}[\cdot]\!=\!(\cdot)^{2/\pi}$,
  $\mathrm{CBRT}[\cdot]\!=\!\sqrt[3]{\cdot}$
\Ensure Ellipse semi-axes $a,b$ Q(32,20) [\si{\milli\meter}]; peak pressure
  $p_{\max}$ Q(32,8) [\si{\mega\pascal}]; ellipticity $k$
\Statex \emph{Every division routes to one shared 80-bit iterative divider
  \textsc{Div} (\texttt{divu}, nine invocations over the chain); both LUTs use
  power-of-two strides so their interpolation collapses to a shift, not a divide.}
\State $R_x \gets \Call{Div}{1\!\ll\!35,\ 2A_{\min}}$
  \Comment{reduced radius, lengthwise}
\State $R_y \gets \Call{Div}{1\!\ll\!35,\ 2B_{\max}}$
  \Comment{reduced radius, profile}
\State $R \gets \Call{Div}{R_x R_y,\ R_x\!+\!R_y}$
  \Comment{effective radius, $1/R\!=\!1/R_x\!+\!1/R_y$}
\State $\alpha \gets \Call{Div}{R_x\!\ll\!16,\ R_y}$
  \Comment{curvature ratio}
\State $k \gets \mathrm{POW}[\alpha]$
  \Comment{$k\!=\!\alpha^{2/\pi}$; LUT $+$ shift interp}
\State $E_e \gets 1 + \Call{Div}{(\pi/2\!-\!1)\!\ll\!s,\ \alpha}$
  \Comment{2nd-kind elliptic-integral approx}
\State $t \gets k^{2}\,E_e\,R\,W$
  \Comment{staged multiply, each partial $\le$ 64 bit (avoids $2^{95}$)}
\State $C_a \gets \Call{Div}{6t\!\ll\!8,\ \pi E'}$
  \Comment{cube-root argument}
\State $a \gets \mathrm{CBRT}[C_a]$
  \Comment{LUT $+$ interval reduction}
\State $b \gets \Call{Div}{a\!\ll\!16,\ k}$
  \Comment{$b\!=\!a/k$: whole chain needs one cube root}
\State $p_{\max} \gets \Call{Div}{3Wk\!\ll\!s,\ 2\pi a^{2}}$
  \Comment{$p_{\max}\!=\!3W/(2\pi ab)$}
\State \Return $a,\ b,\ p_{\max},\ k$
\end{algorithmic}
\end{algorithm}

\section{Implementation and Silicon Bring-Up}
\label{sec:impl}

\subsection{Open-Source Fixed-Point Flow}
The four-stage datapath (Alg.~\ref{alg:gap}--Alg.~\ref{alg:bhseq}) is realized
entirely in fixed-point on a retired data-center accelerator card (Inspur
YPCB-00338) carrying a Xilinx Kintex-7 XC7K480T, through a fully open-source
flow with no vendor tools and no floating-point IP (Fig.~\ref{fig:setup}). LiteX emits each SoC as
CSR-mapped compute cores behind a \texttt{jtagbone} JTAG-to-Wishbone bridge;
\texttt{yosys} synthesizes the RTL, \texttt{nextpnr-xilinx} (openXC7) places and
routes against \texttt{prjxray} device and timing data, \texttt{prjxray} emits
the bitstream, \texttt{openFPGALoader} programs it over JTAG, and the same JTAG
link reads results back to the host~\cite{shah2019yosys,openxc7}. Since every
signal obeys the fixed-point contract of Table~\ref{tab:fixedpoint} and no logic
depends on a proprietary float unit, the silicon result is bit-reproducible and
auditable end to end. Inputs (tooth-surface samples, gap stencil, contact load)
are CSR-written rather than compiled in, so one bitstream evaluates arbitrary
tooth surfaces: changing the gear design is a host register write, not a
place-and-route cycle.

\subsection{Four SoCs, Bit-Exact Silicon}
We built and validated four SoCs; in each, the JTAG-read output matched a golden
software model bit-for-bit. \texttt{contact\_core} implements the Stage~A/B
geometry alone: its \texttt{perp}$^2$-gated argmin tree returned gap code
\num{3622} at contact index~9. \texttt{bhseq\_core} implements the serialized
Stage~D Hertz solver of Alg.~\ref{alg:bhseq}. \texttt{dual\_core} fuses A/B and
D behind a single readback. \texttt{tri\_core} closes the full
A/B\,$+$\,C$\rightarrow$D path: with a gap stencil and contact load streamed in
over Wishbone, on-chip curvature extraction produced relative principal
curvatures ($A_-\!=\!\num{6995}$, $B_-\!=\!\num{20971}$ in Q(32,20)), and the
Hertz stage returned a contact ellipse $a\!=\!\SI{0.909}{\milli\meter}$,
$b\!=\!\SI{0.452}{\milli\meter}$, $p_{\max}\!=\!\SI{1162.5}{\mega\pascal}$,
ellipticity $k\!=\!2.01$---all bit-exact against golden. The complete
\texttt{tri\_core} fits in \num{197377} LUTs (\SI{33}{\percent} of the fabric),
\num{4851} FF, \num{4638} CARRY4, and $2{\times}$RAMB36$+$$1{\times}$RAMB18 with no
DSP, at a synthesized
$f_{\max}$ of \SI{75}{\mega\hertz} (Table~\ref{tab:resource}).

\subsection{Two Bring-Up Pitfalls}
Two hardware issues dominated bring-up, both traceable to the open toolchain
rather than the algorithm. First, \texttt{yosys} inferred the wide datapath
multipliers as DSP48E1 cascades (\texttt{ACOUT}$\rightarrow$\texttt{ACIN}
chaining) that crashed the \texttt{nextpnr} router; synthesizing with
\texttt{-nodsp} remaps every multiply into LUT/CARRY4 fabric, restoring
routability at the cost of inflated LUT usage and DSP${}=0$
(Table~\ref{tab:resource}). Second, the LiteX-generated constraints omitted a
\texttt{create\_clock} on the \SI{50}{\mega\hertz} clock, leaving the
\texttt{perp}$^2$ argmin multiplier chain untimed; it violated setup at
\SI{50}{\mega\hertz} and produced \emph{stable but wrong} silicon. Offline
simulation of the post-synthesis netlist reproduced the correct golden output,
isolating the fault to timing rather than logic; dividing the board clock to
\SI{12.5}{\mega\hertz} restored correct operation, while a properly constrained
pipelined build stays within the \SI{75}{\mega\hertz} $f_{\max}$ of
Table~\ref{tab:resource}.

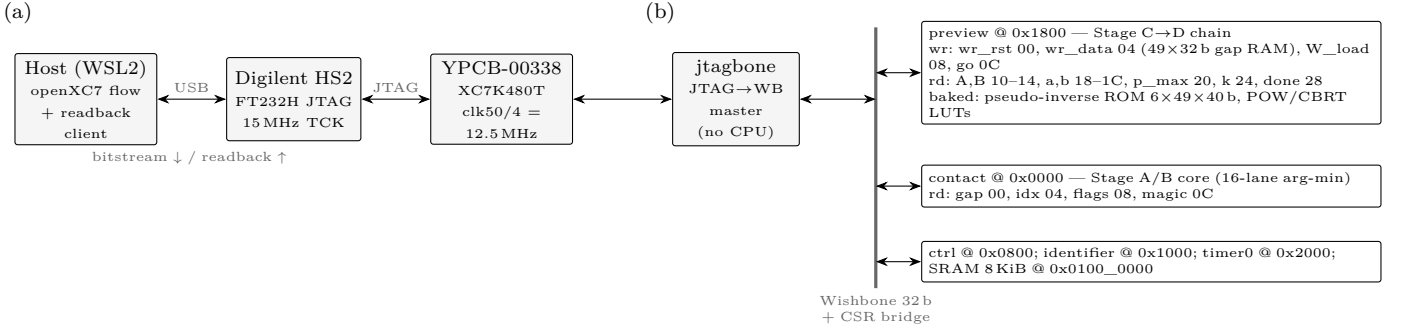
\begin{figure*}[!t]
  \centering
  \begin{tikzpicture}[
    font=\scriptsize,
    >={Stealth[length=1.6mm]},
    box/.style={draw, rounded corners=1pt, align=center, inner sep=2.5pt, fill=black!4},
    csr/.style={draw, rounded corners=1pt, align=left, inner sep=2.5pt, font=\tiny},
    buslbl/.style={font=\tiny, text=black!60, inner sep=1pt},
  ]
    \node[box, text width=17mm, minimum height=11mm] (host) at (0,0)
      {Host (WSL2)\\{\tiny openXC7 flow}\\{\tiny $+$ readback client}};
    \node[box, text width=16mm, minimum height=11mm] (prog) at (2.75,0)
      {Digilent HS2\\{\tiny FT232H JTAG}\\{\tiny \SI{15}{\mega\hertz} TCK}};
    \node[box, text width=17mm, minimum height=11mm] (fpga) at (5.5,0)
      {YPCB-00338\\{\tiny XC7K480T}\\{\tiny clk50/4 $=$ \SI{12.5}{\mega\hertz}}};
    \draw[<->] (host) -- node[buslbl, above=0.5mm]{USB} (prog);
    \draw[<->] (prog) -- node[buslbl, above=0.5mm]{JTAG} (fpga);
    \node[buslbl, anchor=north] at (1.37,-0.65) {bitstream $\downarrow$ / readback $\uparrow$};
    \node[font=\bfseries\footnotesize] at (-0.9,1.15) {(a)};
    \node[font=\bfseries\footnotesize] at (7.6,1.15) {(b)};
    \node[box, text width=15mm, minimum height=9mm] (jb) at (8.6,0)
      {\texttt{jtagbone}\\{\tiny JTAG$\to$WB master}\\{\tiny (no CPU)}};
    \draw[<->] (fpga) -- (jb);
    \draw[line width=1.2pt, black!60] (10.45,0.95) -- (10.45,-2.5)
      node[buslbl, below=0.5mm, align=center]{Wishbone 32\,b\\$+$ CSR bridge};
    \draw[<->] (jb) -- (10.45,0);
    \node[csr, anchor=west, text width=59mm] (prev) at (11.05,0.35)
      {\textbf{\texttt{preview} @ 0x1800} --- Stage C$\to$D chain\\
       wr:\ \texttt{wr\_rst} 00, \texttt{wr\_data} 04 ($49{\times}$32\,b gap RAM),
       \texttt{W\_load} 08, \texttt{go} 0C\\
       rd:\ \texttt{A,B} 10--14, \texttt{a,b} 18--1C, \texttt{p\_max} 20,
       \texttt{k} 24, \texttt{done} 28\\
       baked: pseudo-inverse ROM $6{\times}49{\times}$40\,b, POW/CBRT LUTs};
    \node[csr, anchor=west, text width=59mm] (cont) at (11.05,-1.15)
      {\textbf{\texttt{contact} @ 0x0000} --- Stage A/B core (16-lane arg-min)\\
       rd:\ \texttt{gap} 00, \texttt{idx} 04, \texttt{flags} 08, \texttt{magic} 0C};
    \node[csr, anchor=west, text width=59mm] (misc) at (11.05,-2.15)
      {\texttt{ctrl} @ 0x0800;\ \texttt{identifier} @ 0x1000;\ \texttt{timer0} @ 0x2000;\
       SRAM 8\,KiB @ 0x0100\_0000};
    \draw[<->] (10.45,0.35) -- (prev.west);
    \draw[<->] (10.45,-1.15) -- (cont.west);
    \draw[<->] (10.45,-2.15) -- (misc.west);
  \end{tikzpicture}
  \caption{Silicon-validation platform. (a)~Fully open-source physical chain:
  bitstreams built by openXC7 (yosys$+$nextpnr-xilinx$+$prjxray) are loaded
  through a Digilent HS2 programmer, and results are read back bit-exactly
  over the same JTAG link against a software golden model---no vendor tools,
  no floating-point IP. (b)~The \texttt{tri\_core} SoC as generated by LiteX:
  a \texttt{jtagbone} bridge is the only bus master (no on-chip CPU); the
  Stage~C$\to$D \texttt{preview} core exposes a CSR write port that streams an
  arbitrary $49$-sample gap stencil into on-chip RAM plus a load register and
  a \texttt{go} strobe, so changing the evaluated tooth surface is a register
  write, not a re-synthesis; results ($A_-,B_-,a,b,p_{\max},k$) read back over
  the same bus. Addresses are the generated LiteX map, verified on silicon.}
  \label{fig:setup}
\end{figure*}
\begin{table*}[!t]
\caption{Post-Route Resource Utilization of the Contact-Analysis SoCs on the Xilinx XC7K480T}
\label{tab:resource}
\centering
\footnotesize
\begin{tabular}{l
    S[table-format=6.0]
    S[table-format=6.0]
    S[table-format=4.0]
    c
    S[table-format=4.0]
    S[table-format=2.0]}
\toprule
{SoC} & {LUT (SLICE\_LUTX)} & {FF} & {CARRY4} & {BRAM 36/18} & {DSP48E1} & {$f_{\max}$ (MHz)} \\
\midrule
contact\_core (A/B, $L=16$) &  40000 & {\textemdash} & {\textemdash} &  2/0 &    0 & {\textemdash} \\
bhseq\_core (D, serialized) &  40000 &   857 &  1029 &  0/0 &    0 & 95 \\
tri\_core (A/B$+$C$\rightarrow$D, CSR) & 197377 &  4851 &  4638 &  2/1 &    0 & 75 \\
\midrule
\textbf{XC7K480T total} & 597200 & 597200 & {\textemdash} & 955/1910 & 1920 & {\textemdash} \\
\bottomrule
\end{tabular}

\vspace{2pt}
{\footnotesize\raggedright
All cores were placed and routed with openXC7/nextpnr in \texttt{-nodsp} mode, which
maps multipliers into LUT/CARRY4 fabric to avoid a DSP-cascade router bug; hence
\mbox{DSP48E1 $=0$} and the reported LUT usage is inflated. A production build that
targets the DSP48E1 slices would substantially reduce the LUT count. LUT figures for
contact\_core and bhseq\_core are post-synthesis estimates ($\approx\!40\,000$ each);
tri\_core occupies about \SI{33}{\percent} of the SLICE\_LUTX budget. \par}
\end{table*}

\section{Evaluation}
\label{sec:eval}

All numbers below are measured or computed, not modelled. The four
contact-analysis SoCs were synthesized, placed, and routed with the fully
open-source openXC7 flow (yosys\,$+$\,nextpnr-xilinx\,$+$\,prjxray, no Vivado
and no floating-point IP) and run on a decommissioned data-center card (Inspur
YPCB-00338, Xilinx XC7K480T) over a JTAG (\texttt{jtagbone}) link
\cite{shah2019yosys,openxc7}. Silicon correctness is established by bit-exact
JTAG read-back against a golden software model for every core:
\texttt{contact\_core} (Stage A/B geometry), \texttt{bhseq\_core} (Stage D),
\texttt{dual\_core} (A/B$+$D in one read-back), and \texttt{tri\_core}
(A/B$+$C$\rightarrow$D). As a representative full-chain point,
\texttt{tri\_core} accepts a gap stencil and a load through the CSR interface
and returns on-chip principal curvatures that reduce to contact semi-axes
$a=\SI{0.909}{\milli\meter}$, $b=\SI{0.452}{\milli\meter}$ (ellipticity
$k=a/b=2.01$) and $p_{\max}=\SI{1162.5}{\mega\pascal}$, matching the golden
reference to the last bit. Every result is thus traceable to a specific,
reproducible fixed-point datapath rather than to a trained black box.

The mechanical test article for all accuracy, transmission-error, and
surrogate comparisons is the hypoid pair s0003
(Table~\ref{tab:gearcase}): $7/70$ teeth, \ang{90} shaft angle,
\SI{19}{\milli\meter} offset. Fig.~\ref{fig:gearcase} shows its drive-side
flank sheets in mesh together with the TCA mating point and the
$80{\times}80$ gap-stencil footprint that the on-chip Stage~C consumes---the
same real geometry, not a synthetic benchmark.

\subsection{Accuracy versus full LCP}
\label{sec:eval:accuracy}
The physics-direct preview is a bulk-Hertz estimator, and it should be read as
an engineering-magnitude quantity, not a full LCP solve. Evaluated on case
s0003 at the design mesh phase $\Theta=\SI{540}{\micro\radian}$
($111.4''$), against a ground truth of full Boussinesq influence-coefficient
LCP \cite{boussinesq1885}, the preview peak pressure lands squarely in the LCP
body-Hertz band (mean to $p_{90}$) and scales strictly as $W^{1/3}$
(Fig.~\ref{fig:accuracy}). Across the four operating torques the mean deviation
against $p_{90}$ is $-22\%$ (per-torque $-13\%$ to $-31\%$; Table~\ref{tab:accuracy}).
The gap the preview cannot close is physical, not numerical: the absolute LCP
peak is a tooth-tip edge-load spike reaching $\approx\!5\times$ the body
pressure, which is governed by half-space edge kinematics and is only
recovered by the LCP. Fig.~\ref{fig:pressure} makes this split spatially
explicit at $T=\SI{64.5}{\newton\meter}$: the on-chip Hamrock--Brewe
footprint lies on the LCP contact band with matching position, orientation,
and bulk pressure level, while the $>\!\SI{5}{\giga\pascal}$ cells lining
the tooth-tip edge of the band are the tip-interference spikes only the LCP
resolves. This delineates the
applicability domain cleanly---bulk Hertz for fast, deterministic preview;
full LCP where the edge spike drives the design margin.

\subsection{Physics-direct versus neural surrogate}
\label{sec:eval:nn}
A closed-form Hertz/Hamrock--Brewe evaluation extrapolates in load where a
neural surrogate cannot, at zero training cost (Fig.~\ref{fig:nn}). On a
synthetic $(A_-,B_-,W)$ domain an MLP reaches \SI{0.94}{\percent} RMSE inside
its training box but degrades to \SI{27.9}{\percent} once the load is pushed to
$3$--$\SI{9}{\kilo\newton}$---a $30\times$ collapse---whereas the physical
evaluation is exact by construction ($\approx\!0$) because it carries no fitted
parameters. Anchored instead to measured LCP $p_{90}$ (case s0003, load
extrapolation), the surrogate is nominally perfect on its trained loads
(\SI{0.0}{\percent}) but reaches \SI{20.3}{\percent} outside them, while the
physical preview holds a stable, physically-explained offset
(\SI{29.2}{\percent} trained region, \SI{18.9}{\percent} extrapolation region).
The decisive comparison is on the extrapolation region: the zero-training
physical model (\SI{18.9}{\percent}) is more accurate there than the
\emph{trained} surrogate (\SI{20.3}{\percent}). The surrogate's cost is also
non-trivial and recurring: each training sample is one full LCP solve
($\approx\SI{2}{\second}$), so a 400-sample set costs $\approx\SI{13}{\minute}$
and still offers no extrapolation guarantee. The physical path needs no data,
extrapolates by Hertzian construction, and stays interpretable and
fixed-point auditable end-to-end, since curvature, ellipse, and pressure are
all physical quantities.

\subsection{Determinism and latency}
\label{sec:eval:latency}
Because the datapath has no data-dependent branch, every stage completes in a
fixed cycle count with zero jitter ($\sigma=0$; Table~\ref{tab:latency},
Fig.~\ref{fig:bhseq}(b)). The Stage A/B gap kernel---a pose transform, a $3$D
gap, and a $\text{perp}^2$-gated \textsc{argmin} tree (Alg.~\ref{alg:gap})---
resolves in $26$ cycles; the combinational Stage\,C$+$D preview in $58$; the
serialized Hamrock--Brewe Stage D (Alg.~\ref{alg:bhseq}) in $756$; and the full
C$\rightarrow$D chain in $806$ cycles, i.e.\ \SI{64}{\micro\second} on the
\SI{12.5}{\mega\hertz} board build and \SI{16}{\micro\second} at the
\SI{50}{\mega\hertz} pipelined target. A single \texttt{go} pulse triggers the
FSM and \texttt{done} asserts at a constant offset. The contrast with software
is not merely speed but predictability: the CPU preview runs at
\SI{277}{\micro\second} with \SI{7}{\percent} jitter and the full LCP solve at
\SI{2005}{\milli\second}, whereas the FPGA latency is a compile-time constant---
the property a hard-real-time digital twin actually requires.

\subsection{Resource and throughput scaling}
\label{sec:eval:resource}
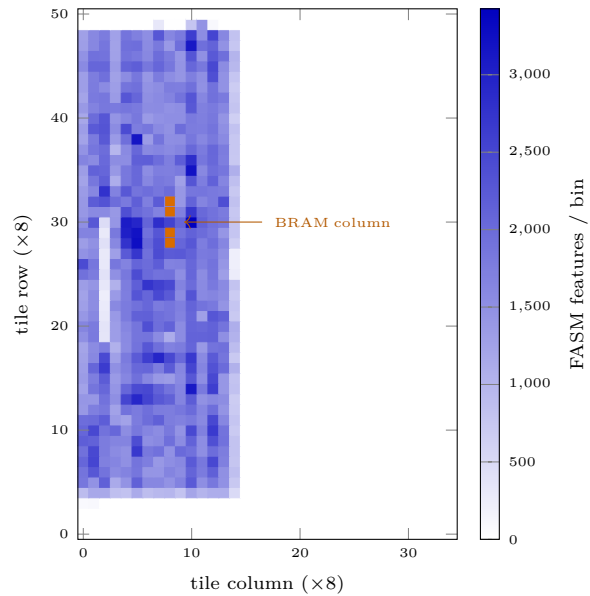
\begin{figure}[!t]
  \centering
  \begin{tikzpicture}
    \begin{axis}[
      width=6.6cm, height=8.6cm,
      xlabel={tile column ($\times 8$)}, ylabel={tile row ($\times 8$)},
      xlabel style={font=\scriptsize}, ylabel style={font=\scriptsize},
      ticklabel style={font=\tiny},
      xmin=-0.5, xmax=34.5, ymin=-0.5, ymax=50.5,
      colormap={whiteblue}{color(0)=(white) color(1)=(blue!75!black)},
      colorbar, colorbar style={width=2.5mm, ticklabel style={font=\tiny},
        ylabel={FASM features / bin}, ylabel style={font=\scriptsize}},
      point meta max=3426,
      axis on top,
    ]
      \addplot[matrix plot*, mesh/cols=35, point meta=explicit]
        table[x=x, y=y, meta=z, col sep=comma] {data/floorplan.csv};
      \addplot[only marks, mark=square*, mark size=1.6pt, orange!85!black]
        coordinates {(8,28) (8,29) (8,31) (8,32)};
      \draw[->, orange!70!black, thin] (axis cs:16.5,30) -- (axis cs:9.3,30);
      \node[font=\tiny, anchor=west, text=orange!70!black] at (axis cs:16.8,30)
        {BRAM column};
    \end{axis}
  \end{tikzpicture}
  \caption{Measured \texttt{tri\_core} die occupancy on the XC7K480T, parsed
  from the final FASM bitstream database ($8{\times}8$-tile bins; color $=$
  configured-feature count, a proxy for logic density). The \texttt{-nodsp}
  flow spreads all multipliers across LUT/CARRY4 fabric---\num{28573} CLB
  tiles carry configuration (\num{197377} LUTs, \SI{33}{\percent}), every DSP
  column is untouched ($\mathrm{DSP}{=}0$), and the single occupied block-RAM
  column segment (orange; $2{\times}$RAMB36 $+$ $1{\times}$RAMB18) holds the
  CSR gap RAM and the baked pseudo-inverse/LUT ROMs.}
  \label{fig:floorplan}
\end{figure}

The design fits comfortably and its scaling ceiling is now measured, not
conjectured. The full \texttt{tri\_core} occupies \num{197377} LUTs
($\approx\!33\%$ of the SLICE\_LUTX budget), \num{4851} FF, \num{4638} CARRY4,
$2{\times}$RAMB36$+$$1{\times}$RAMB18, and $0$ DSP at a system $f_{\max}$ of \SI{75}{\mega\hertz}
(Table~\ref{tab:resource}); the measured die occupancy parsed from the final
bitstream database is shown in Fig.~\ref{fig:floorplan}. The zero DSP count and
inflated LUT usage are an
artifact of the \texttt{-nodsp} mode, which maps every multiplier into
LUT/CARRY4 fabric to dodge a DSP-cascade routing bug in nextpnr; a
DSP-targeted production build would cut the LUT count substantially. Sweeping
the geometric gap core over $L=8/16/32/64$ compute lanes (yosys) gives
$258/498/978/\num{1938}$ DSP and $4207/8132/\num{16000}/\num{30443}$ LC, i.e.\ a
clean $\text{DSP}\approx 30L+18$ (Fig.~\ref{fig:throughput}). At $L=64$ the
$\num{1938}$-DSP demand exceeds the XC7K480T's $\num{1920}$-DSP ceiling, so the
practical limit is $L\approx63$ and the accelerator is \emph{DSP-bound}---
correcting our earlier LUT/timing-bound conjecture and setting the correct
knob for a larger part.

\subsection{Loaded transmission error}
\label{sec:eval:te}
Loaded TE is the primary gear NVH excitation \cite{te2018nvh,evgear2025nvh},
and the preview reproduces its waveform while under-predicting only its
amplitude. Along the mesh phase the preview and the full LCP share the
single-to-double tooth-pair transition and the overall LTE shape, because both
consume the same kinematic contact path (Fig.~\ref{fig:te}); they differ only
in contact compliance, $c_{\text{LCP}}=\SI{8.69e-6}{\milli\meter\per\newton}$
versus $c_{\text{pre}}=\SI{6.62e-6}{\milli\meter\per\newton}$ ($0.76\times$, the
Hertzian model being stiffer than Boussinesq). The resulting loaded-TE
peak-to-peak, the quantity that drives whine, is $1.61''$ (LCP) versus $1.31''$
(preview), a $-19\%$ difference, against an unloaded $\text{TE}_0$ peak-to-peak
of $3.40''$. The preview therefore places the NVH-relevant order and phase
correctly at a bounded, characterized amplitude bias.

\subsection{Multi-tooth load sharing}
\label{sec:eval:multitooth}
Beyond the single-pair patch, a full multi-tooth solve---per-tooth conjugate
separation fields under one shared rigid-body rotation, block-diagonal
structural compliance from a shell Rayleigh--Ritz model of both members, and
a torque-balance LCP---engages three tooth pairs at
$T=\SI{64.5}{\newton\meter}$: the reference pair carries \SI{74.6}{\percent}
(\SI{1521}{\newton}), its neighbors \SI{22.6}{\percent} and
\SI{2.8}{\percent} (Fig.~\ref{fig:multitooth}). The third pair sits at the
engagement threshold, so the engaged count (and with it the regularized
peak) is sensitive to small phase or numerical perturbations; the robust
outputs are the sharing gradient and the mesh rotation. Two cross-checks
fall out.
First, structural compliance is now visible: the loaded mesh rotation grows
to $124.4''$ versus $111.4''$ for the rigid-tooth Boussinesq patch, a
$+12\%$ tooth-bending contribution. Second, the two solvers regularize the
same physical tooth edge differently: the multi-tooth pipeline trims
boundary-inadmissible edge cells and reports an incomplete-ellipse
line-contact peak ($\mathrm{CP}_{\max}=\SI{2714}{\mega\pascal}$), so the
tip-edge spikes of Fig.~\ref{fig:pressure} are excluded from its headline
metric (individual cells still reach \SI{5.8}{\giga\pascal} at the incipient
$k{=}{+}1$ contact)---a reminder that the certification-grade edge number is
a modeling choice of the host solver, while the on-chip preview
(\SI{2038}{\mega\pascal}) consistently tracks the bulk level under both
treatments.

\subsection{Fixed-point precision}
\label{sec:eval:precision}
The geometric gap does not need double precision. Under the fully fixed-point
contract (Table~\ref{tab:fixedpoint}: coordinates Q(24,16)\,mm, normals Q1.19,
curvatures Q(32,20), $p_{\max}$ Q(32,8)), the gap error falls monotonically
with coordinate wordwidth from \SI{3.3}{\micro\meter} at $16$ bit to
\SI{0.011}{\micro\meter} at $24$ bit, crossing sub-micron at $\geq\!20$ bit,
while the normal-vector wordwidth saturates at $\geq\!18$ bit
(Fig.~\ref{fig:precision}). Twenty-bit coordinates thus deliver sub-micron gaps
at a fraction of an fp64 footprint, which is what makes the all-fixed-point,
DSP-lean datapath above both auditable and area-efficient.

This scaling is not empirical accident but follows from a first-order
quantization bound. With $c_f$ fractional coordinate bits and $r_f$ fractional
normal bits, the three independent error sources of Eq.~\eqref{eq:gaplane} are
the quantized difference vector ($\lVert\boldsymbol{\varepsilon}_d
\rVert_\infty\!\le\!2^{-c_f}$, projected through a unit normal with
$\lVert\mathbf{n}\rVert_1\!\le\!\sqrt{3}$), the quantized normal acting on the
true offset ($|\Delta\mathbf{n}\cdot\mathbf{d}|\!\le\!3\cdot
2^{-(r_f+1)}\lVert\mathbf{d}\rVert_\infty$), and the single output rounding, so
\begin{equation}
  |\varepsilon_g| \;\le\; \sqrt{3}\,2^{-c_f}
  \;+\; 3\,\lVert\mathbf{d}\rVert_\infty 2^{-(r_f+1)} \;+\; 2^{-(c_f+1)}.
  \label{eq:qbound}
\end{equation}
The first term dominates and predicts one-LSB-proportional decay: the measured
sweep stays within $0.3$--$0.9$ LSB of the coordinate format at every width
(e.g.\ bound \SI{8.7}{\micro\meter} vs.\ measured \SI{3.3}{\micro\meter} at
$c_f{=}8$; bound \SI{0.034}{\micro\meter} vs.\ \SI{0.011}{\micro\meter} at
$c_f{=}16$). The bound also locates the normal-width saturation point:
with $\lVert\mathbf{d}\rVert_\infty$ of order \SI{1}{\milli\meter} near
contact, the second term falls below the first at $r_f\!\gtrsim\!17$, exactly
where the measured curve flattens. Wordlength choices in
Table~\ref{tab:fixedpoint} are therefore certified by
Eq.~\eqref{eq:qbound} rather than tuned by trial.

\begin{table}[!t]
\caption{Evaluated hypoid pair (case s0003) and analysis setup.}
\label{tab:gearcase}
\centering
\footnotesize
\begin{tabular}{ll}
\toprule
Tooth numbers $z_1/z_2$ (ratio) & $7/70$ ($10{:}1$) \\
Shaft angle / hypoid offset & \ang{90} / \SI{19}{\milli\meter} \\
Spiral hand & right \\
Gear mean cone distance & \SI{40.68}{\milli\meter} \\
Pitch angle (gear / pinion) & \ang{83.15} / \ang{6.07} \\
Assembly errors $(E,P,G,\alpha)$ & $(0,0,0,0)$ \\
Material (both members) & steel, $E{=}\SI{209}{\giga\pascal}$, $\nu{=}0.3$ \\
Flank sampling & $16{\times}22$ flank grid; $120{\times}150$ dense \\
Gap stencil (Stage C input) & $80{\times}80$ tangent-plane field \\
Operating torques & $32.25$--\SI{258}{\newton\meter} (Table~\ref{tab:accuracy}) \\
Design mesh phase & $\Theta_0=\SI{540}{\micro\radian}$ ($111.4''$) \\
\bottomrule
\end{tabular}
\end{table}

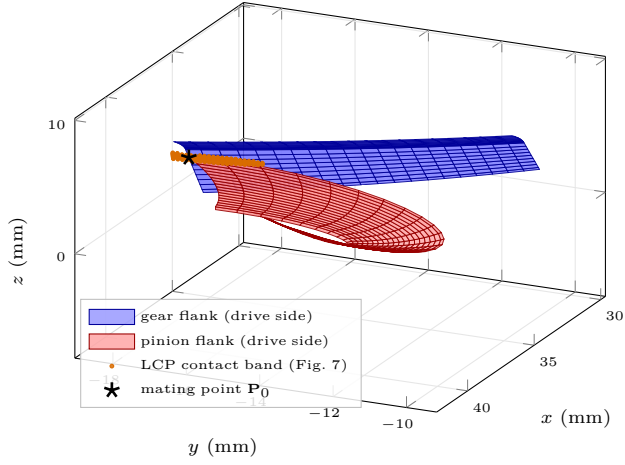
\begin{figure}[!t]
  \centering
  \begin{tikzpicture}
    \begin{axis}[
      width=8.6cm, height=7.0cm,
      view={115}{28},
      xlabel={$x$ (\si{\milli\meter})}, ylabel={$y$ (\si{\milli\meter})},
      zlabel={$z$ (\si{\milli\meter})},
      label style={font=\scriptsize}, ticklabel style={font=\tiny},
      grid=major, major grid style={gray!20},
      legend style={font=\tiny, at={(0.01,0.01)}, anchor=south west,
                    draw=gray!50, fill=white, fill opacity=0.85, text opacity=1},
      legend cell align=left,
      z buffer=sort,
    ]
      \addplot3[surf, shader=faceted, mesh/cols=20,
                colormap={gb}{color(0)=(blue!25) color(1)=(blue!55)},
                faceted color=blue!60!black, line width=0.05pt, opacity=0.9,
                forget plot]
        table[x=x, y=y, z=z, col sep=comma] {data/flank_gear.csv};
      \addlegendimage{area legend, fill=blue!40, draw=blue!60!black}
      \addlegendentry{gear flank (drive side)}
      \addplot3[surf, shader=faceted, mesh/cols=20,
                colormap={pr}{color(0)=(red!20) color(1)=(red!50)},
                faceted color=red!60!black, line width=0.05pt, opacity=0.9,
                forget plot]
        table[x=x, y=y, z=z, col sep=comma] {data/flank_pinion.csv};
      \addlegendimage{area legend, fill=red!35, draw=red!60!black}
      \addlegendentry{pinion flank (drive side)}
      \addplot3[only marks, mark=*, mark size=0.7pt, orange!85!black]
        table[x=x, y=y, z=z, col sep=comma] {data/flank_band.csv};
      \addlegendentry{LCP contact band (Fig.~\ref{fig:pressure})}
      \addplot3[only marks, mark=star, mark size=3pt, line width=1pt, black]
        coordinates {(41.679,-16.164,7.949)};
      \addlegendentry{mating point $\mathbf{P}_0$}
    \end{axis}
  \end{tikzpicture}
  \caption{Case s0003 drive-side tooth flanks in mesh (real geometry, meshing
  frame): the gear and pinion contact-zone flank sheets, the TCA mating point
  $\mathbf{P}_0$ (\SI{9}{\micro\meter} off the dense flank grid), and the LCP
  contact band of Fig.~\ref{fig:pressure} mapped onto the tangent plane at
  $\mathbf{P}_0$ (sag $<\SI{0.1}{\milli\meter}$; cells falling outside the
  extracted flank window are omitted). Stage C samples its $80{\times}80$
  gap stencil on this tangent plane around $\mathbf{P}_0$. The gear axis is
  $z$ through the origin and the pinion axis is $x$ through
  $(0,-19,0)$\,\si{\milli\meter}---the \SI{19}{\milli\meter} hypoid offset of
  Table~\ref{tab:gearcase}.}
  \label{fig:gearcase}
\end{figure}
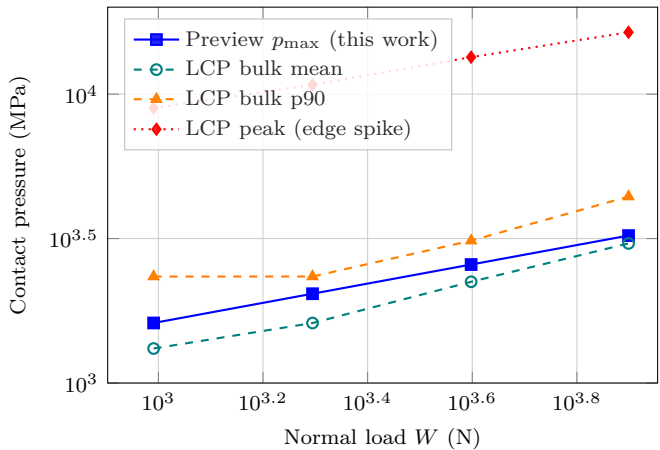
\begin{figure}[!t]
  \centering
  \begin{tikzpicture}
    \begin{axis}[
      width=\columnwidth,
      height=0.74\columnwidth,
      xmode=log, ymode=log,
      log basis x=10, log basis y=10,
      xlabel={Normal load $W$~(\si{\newton})},
      ylabel={Contact pressure~(\si{\mega\pascal})},
      xmin=800, xmax=9000,
      ymin=1000, ymax=20000,
      grid=both,
      major grid style={line width=0.2pt, draw=gray!40},
      minor grid style={draw=gray!15},
      legend pos=north west,
      legend cell align=left,
      legend style={font=\footnotesize, fill opacity=0.85, draw=gray!50},
      tick label style={font=\footnotesize},
      label style={font=\footnotesize},
      mark options={solid},
    ]
      \addplot[thick, solid, color=blue, mark=square*]
        table[x=W, y=preview_pmax, col sep=comma]{data/accuracy.csv};
      \addlegendentry{Preview $p_{\max}$ (this work)}

      \addplot[thick, dashed, color=teal, mark=o]
        table[x=W, y=lcp_mean, col sep=comma]{data/accuracy.csv};
      \addlegendentry{LCP bulk mean}

      \addplot[thick, dashed, color=orange, mark=triangle*]
        table[x=W, y=lcp_p90, col sep=comma]{data/accuracy.csv};
      \addlegendentry{LCP bulk p90}

      \addplot[thick, dotted, color=red, mark=diamond*]
        table[x=W, y=lcp_pmax, col sep=comma]{data/accuracy.csv};
      \addlegendentry{LCP peak (edge spike)}
    \end{axis}
  \end{tikzpicture}
  \caption{Preview bulk pressure versus loaded tooth-contact (LCP) statistics over
  normal load. The fixed-point preview peak (this work) tracks the LCP bulk-Hertz
  band (mean to p90) within about \SI{22}{\percent} and scales as $W^{1/3}$, whereas
  the LCP edge-load peak is roughly $5\times$ higher; this delineates the
  applicability domain: bulk Hertz for fast preview, full LCP for edge-spike design.}
  \label{fig:accuracy}
\end{figure}
\begin{table*}[!t]
\caption{Fixed-point combined-Hertz preview versus full LCP contact pressure. The preview tracks the body Hertz level ($\sim$mean--$p_{90}$), averaging $-22\%$ against $p_{90}$; tooth-edge load peaks ($p_{\text{peak}}\!\approx\!5\times$ body) still require the full LCP solve.}
\label{tab:accuracy}
\centering
\footnotesize
\begin{tabular}{
  S[table-format=3.2]
  S[table-format=4.0]
  S[table-format=4.0]
  S[table-format=4.0]
  S[table-format=4.0]
  S[table-format=-2.1]
  S[table-format=5.0]
}
\toprule
{$T$ (\si{\newton\meter})} & {$W$ (\si{\newton})} & {$p_{\max}^{\text{prev}}$ (\si{\mega\pascal})} & {$\bar{p}_{\text{LCP}}$ (\si{\mega\pascal})} & {$p_{90}$ (\si{\mega\pascal})} & {err vs $p_{90}$ (\si{\percent})} & {$p_{\text{peak}}$ (edge, \si{\mega\pascal})} \\
\midrule
 32.25 &  979 & 1614 & 1317 & 2336 & -30.9 &  8951 \\
 64.50 & 1971 & 2038 & 1613 & 2337 & -12.8 & 10772 \\
129.00 & 3961 & 2571 & 2241 & 3110 & -17.3 & 13426 \\
258.00 & 7919 & 3239 & 3042 & 4418 & -26.7 & 16372 \\
\bottomrule
\end{tabular}
\end{table*}

\begin{figure}[!t]
  \centering
  \begin{tikzpicture}
    \begin{axis}[
      width=7.5cm, height=4.0cm,
      xlabel={$s$ along flank (\si{\milli\meter})},
      ylabel={$t$ (\si{\milli\meter})},
      label style={font=\scriptsize}, ticklabel style={font=\tiny},
      xmin=-3.4, xmax=3.0, ymin=-1.05, ymax=1.05,
      axis equal image,
      colormap={pmap}{color(0)=(blue!10) color(0.5)=(orange!70) color(1)=(red!85!black)},
      colorbar, colorbar style={width=2.5mm, ticklabel style={font=\tiny},
        ylabel={$p$ (\si{\mega\pascal})}, ylabel style={font=\scriptsize}},
      point meta min=0, point meta max=4000,
      legend style={font=\tiny, at={(0.985,0.03)}, anchor=south east,
                    draw=gray!50, fill=white, fill opacity=0.85, text opacity=1},
      legend cell align=left,
    ]
      \addplot[scatter, only marks, mark=square*, mark size=1.9pt,
               scatter src=explicit, forget plot]
        table[x=s, y=t, meta=p, col sep=comma] {data/pressure_lcp.csv};
      \addplot[black, thick, dashed]
        table[x=s, y=t, col sep=comma] {data/ellipse_preview.csv};
      \addlegendentry{preview Hertz ellipse}
    \end{axis}
  \end{tikzpicture}
  \caption{Spatial view of the accuracy claim at $T=\SI{64.5}{\newton\meter}$
  ($W=\SI{1971}{\newton}$): the full Boussinesq$+$LCP pressure map over the
  contact band ($196$ cells on the adaptive, non-uniformly spaced
  tangent-plane grid; only contacting cells shown; color clipped at
  \SI{4}{\giga\pascal}). The saturated cells concentrated along the
  tooth-tip line $t\!\approx\!\SI{0.5}{\milli\meter}$ are the
  $\approx\!5\times$ tip-edge-loading spikes of Table~\ref{tab:accuracy}
  (12 cells, up to \SI{10.8}{\giga\pascal})---contact truncated by the gear
  tip, a half-space edge effect. Overlaid,
  the physics-direct Hamrock--Brewe footprint computed on-chip
  ($2a=\SI{4.8}{\milli\meter}$, $2b=\SI{0.38}{\milli\meter}$,
  $p_{\max}^{\mathrm{prev}}=\SI{2038}{\mega\pascal}$ vs.\ LCP
  $p_{90}=\SI{2337}{\mega\pascal}$). The preview ellipse tracks the bulk
  band's position, orientation ($-13.8^{\circ}$ from the flank direction),
  and pressure level, while the edge spikes remain LCP-only---the
  applicability split of Sec.~\ref{sec:discussion}.}
  \label{fig:pressure}
\end{figure}
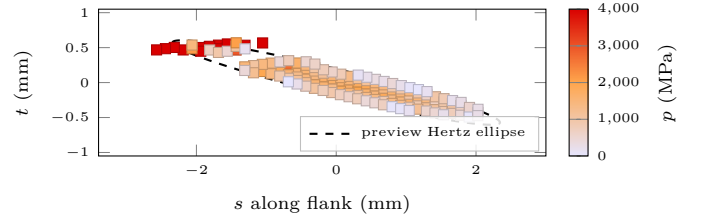

\begin{figure*}[!t]
  \centering
  \begin{tikzpicture}
    \begin{axis}[
      width=0.49\linewidth, height=5.6cm,
      xlabel={Contact load $W$ (\si{\newton})},
      ylabel={Preview RMSE (\si{\percent})},
      xmin=300, xmax=9200, ymin=0, ymax=47,
      title={(a) Synthetic sweep},
      grid=major, grid style={gray!20},
      legend pos=north west, legend cell align=left,
      legend style={font=\footnotesize, draw=none, fill=none},
      title style={font=\footnotesize},
      tick label style={font=\footnotesize},
      label style={font=\footnotesize},
    ]
      \addplot[blue, mark=*, mark size=1.2pt, thick]
        table[x=W, y=nn_rmse, col sep=comma]{data/nn_synth.csv};
      \addlegendentry{NN preview}
      \addplot[red, mark=square*, mark size=1.2pt, thick]
        table[x=W, y=phys_rmse, col sep=comma]{data/nn_synth.csv};
      \addlegendentry{Physical direct}
      \draw[black!70, dashed, thick] (axis cs:3000,0) -- (axis cs:3000,47);
      \node[anchor=south east, font=\scriptsize, text=black!55]
        at (axis cs:2900,3) {train box};
      \node[anchor=south west, font=\scriptsize, text=black!55]
        at (axis cs:3120,3) {extrapolation};
    \end{axis}
  \end{tikzpicture}
  \hfill
  \begin{tikzpicture}
    \begin{axis}[
      width=0.49\linewidth, height=5.6cm,
      xlabel={Contact load $W$ (\si{\newton})},
      ylabel={Peak pressure $p_{\max}$ (\si{\mega\pascal})},
      xmin=400, xmax=7300, ymin=1200, ymax=4300,
      title={(b) Real LCP anchoring (s0003)},
      grid=major, grid style={gray!20},
      legend pos=north west, legend cell align=left,
      legend style={font=\footnotesize, draw=none, fill=none},
      title style={font=\footnotesize},
      tick label style={font=\footnotesize},
      label style={font=\footnotesize},
    ]
      \addplot[black, mark=*, mark size=1.2pt, thick]
        table[x=W, y=lcp_p90, col sep=comma]{data/nn_extrap.csv};
      \addlegendentry{LCP $p_{90}$ (truth)}
      \addplot[blue, mark=triangle*, mark size=1.5pt, thick]
        table[x=W, y=nn_pmax, col sep=comma]{data/nn_extrap.csv};
      \addlegendentry{NN preview}
      \addplot[red, mark=square*, mark size=1.2pt, thick]
        table[x=W, y=phys_pmax, col sep=comma]{data/nn_extrap.csv};
      \addlegendentry{Physical direct}
      \draw[black!70, dashed, thick] (axis cs:3065,1200) -- (axis cs:3065,4300);
      \node[anchor=south east, font=\scriptsize, text=black!55]
        at (axis cs:2970,1310) {train};
      \node[anchor=south west, font=\scriptsize, text=black!55]
        at (axis cs:3160,1310) {extrapolation};
    \end{axis}
  \end{tikzpicture}
  \caption{Load-extrapolation of the closed-form Hertz/Hamrock--Brewe physical
    preview versus a neural surrogate. (a) Synthetic sweep: the NN tracks the
    truth inside its training box ($W\le\SI{3000}{\newton}$, dashed line) but its
    RMSE collapses to \SI{27.9}{\percent} once extrapolated, whereas the
    zero-training physical-direct evaluation stays at $\approx\!0$. (b) Anchored
    to measured LCP $p_{90}$ peak pressures (case s0003): outside the trained
    loads the NN error reaches \SI{20.3}{\percent}, while the physical preview
    extrapolates at \SI{18.9}{\percent}, i.e.\ below the trained surrogate with no
    retraining.}
  \label{fig:nn}
\end{figure*}
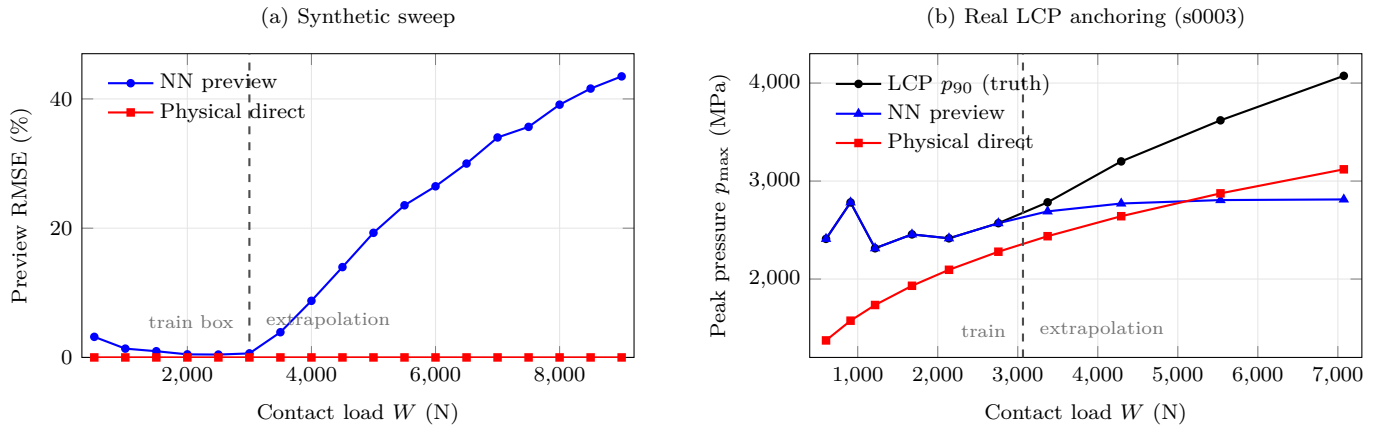
\begin{table}[!t]
\caption{Deterministic bounded latency (zero jitter).}
\label{tab:latency}
\centering
\footnotesize
\setlength{\tabcolsep}{4pt}
\begin{tabular}{l S[table-format=3.0] S[table-format=2.1] S[table-format=2.2]}
\toprule
 & & \multicolumn{2}{c}{Latency (\si{\micro\second})} \\
\cmidrule(lr){3-4}
{Core} & {Cycles} & {@\,\SI{12.5}{\mega\hertz}} & {@\,\SI{50}{\mega\hertz}} \\
\midrule
Stage A/B gap kernel                       & 26  & 2.1 & 0.52 \\
Stage C+D preview (comb.)    & 58  & 4.6 & 1.2  \\
Stage D H.--Brewe (serial.)        & 756 & 60  & 15   \\
Full chain C$\to$D  & 806 & 64  & 16   \\
\bottomrule
\end{tabular}
\par\smallskip
{\footnotesize\raggedright Cycle counts are RTL-exact ($\sigma\!=\!0$). CPU
reference: preview \SI{277}{\micro\second} (\SI{7}{\percent} jitter); full LCP
\SI{2005}{\milli\second}.\par}
\end{table}

\begin{figure}[!t]
  \centering
  \begin{tikzpicture}
    \begin{axis}[
      scale only axis,
      width=0.74\columnwidth,
      height=0.50\columnwidth,
      xlabel={Compute lanes $L$},
      ylabel={DSP48E1 slices},
      xmin=0, xmax=68,
      ymin=0, ymax=2100,
      xtick={8,16,32,64},
      axis y line*=left,
      axis x line*=bottom,
      ylabel near ticks,
      tick label style={font=\footnotesize},
      label style={font=\footnotesize},
      legend style={font=\footnotesize, at={(0.02,0.88)}, anchor=north west,
                    draw=none, fill=none, row sep=-1pt},
    ]
      \addplot[only marks, mark=*, mark size=1.6pt, black]
        table[x=L, y=dsp, col sep=comma]{data/throughput.csv};
      \addlegendentry{DSP (measured)}
      \addplot[dashed, thick, black, domain=0:68, samples=2] {30*x+18};
      \addlegendentry{$\mathrm{DSP}\approx 30L+18$}
      \addplot[red, thick, domain=0:68, samples=2] {1920};
      \addlegendentry{1920 DSP ceiling}
      \addlegendimage{only marks, mark=square, blue}
      \addlegendentry{Logic cells (right)}
      \addplot[dotted, gray] coordinates {(63.4,0) (63.4,1920)};
      \node[anchor=south east, font=\footnotesize] at (axis cs:63,1920) {$L\!\approx\!63$};
    \end{axis}
    \begin{axis}[
      scale only axis,
      width=0.74\columnwidth,
      height=0.50\columnwidth,
      xmin=0, xmax=68,
      ymin=0, ymax=32000,
      axis y line*=right,
      axis x line=none,
      ylabel={Logic cells},
      ylabel near ticks,
      y tick label style={/pgf/number format/1000 sep={\,}, font=\footnotesize},
      label style={font=\footnotesize},
      every axis y label/.append style={blue},
      y axis line style={blue},
      y tick style={blue},
    ]
      \addplot[only marks, mark=square, mark size=1.6pt, blue]
        table[x=L, y=lc, col sep=comma]{data/throughput.csv};
    \end{axis}
  \end{tikzpicture}
  \caption{Resource scaling of the geometric gap core versus compute lanes $L$
    (yosys synthesis): DSP48E1 usage grows linearly as $\mathrm{DSP}\approx 30L+18$
    and saturates the XC7K480T ceiling of \num{1920} DSP at $L\approx 63$, showing
    the accelerator is DSP-bound rather than LUT-bound.}
  \label{fig:throughput}
\end{figure}
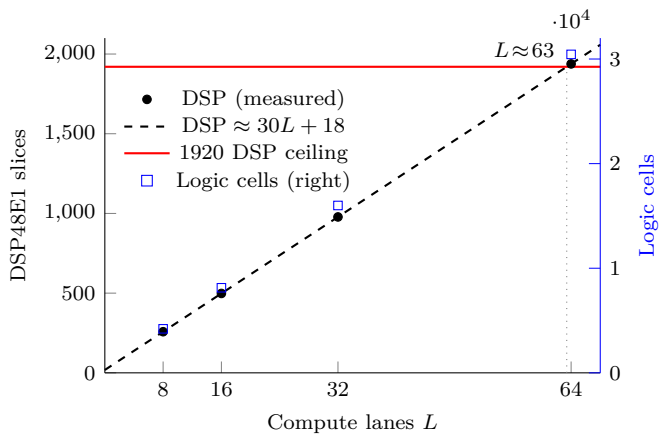
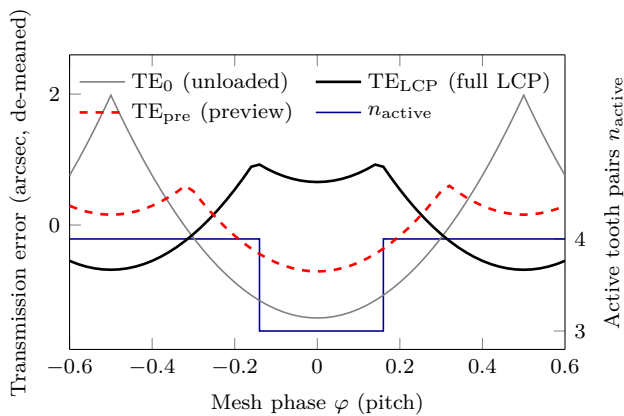
\begin{figure}[!t]
  \centering
  \begin{tikzpicture}
    \begin{axis}[
        scale only axis,
        width=0.74\columnwidth, height=0.44\columnwidth,
        xmin=-0.6, xmax=0.6,
        ymin=2.8, ymax=6.0,
        enlargelimits=false,
        axis y line*=right,
        axis x line=none,
        ytick={3,4},
        ylabel={Active tooth pairs $n_{\mathrm{active}}$},
        tick label style={font=\footnotesize},
        label style={font=\footnotesize},
      ]
      \addplot[blue!55!black, const plot, semithick]
        table[x=phi, y=nactive, col sep=comma]{data/te_curve.csv};
    \end{axis}
    \begin{axis}[
        scale only axis,
        width=0.74\columnwidth, height=0.44\columnwidth,
        xmin=-0.6, xmax=0.6,
        ymin=-1.9, ymax=2.6,
        enlargelimits=false,
        axis y line*=left,
        axis background/.style={fill=none},
        xlabel={Mesh phase $\varphi$ (pitch)},
        ylabel={Transmission error (arcsec, de-meaned)},
        tick label style={font=\footnotesize},
        label style={font=\footnotesize},
        legend columns=2,
        legend style={font=\footnotesize, at={(0.5,0.98)}, anchor=north,
                      draw=none, fill=none,
                      /tikz/every even column/.append style={column sep=6pt}},
        legend cell align=left,
      ]
      \addplot[gray, line width=0.6pt]
        table[x=phi, y=te0, col sep=comma]{data/te_curve.csv};
      \addlegendentry{$\mathrm{TE}_0$ (unloaded)}
      \addplot[black, line width=1.0pt]
        table[x=phi, y=te_lcp, col sep=comma]{data/te_curve.csv};
      \addlegendentry{$\mathrm{TE}_{\mathrm{LCP}}$ (full LCP)}
      \addplot[red, dashed, line width=1.0pt]
        table[x=phi, y=te_pre, col sep=comma]{data/te_curve.csv};
      \addlegendentry{$\mathrm{TE}_{\mathrm{pre}}$ (preview)}
      \addlegendimage{blue!55!black, semithick}
      \addlegendentry{$n_{\mathrm{active}}$}
    \end{axis}
  \end{tikzpicture}
  \caption{Loaded transmission error along the mesh phase. The physics-direct
  preview and the full LCP share the single/double tooth-pair transition and
  the overall waveform; their peak-to-peak values differ by $19\%$, reflecting
  the Hertzian versus Boussinesq contact-compliance models.}
  \label{fig:te}
\end{figure}
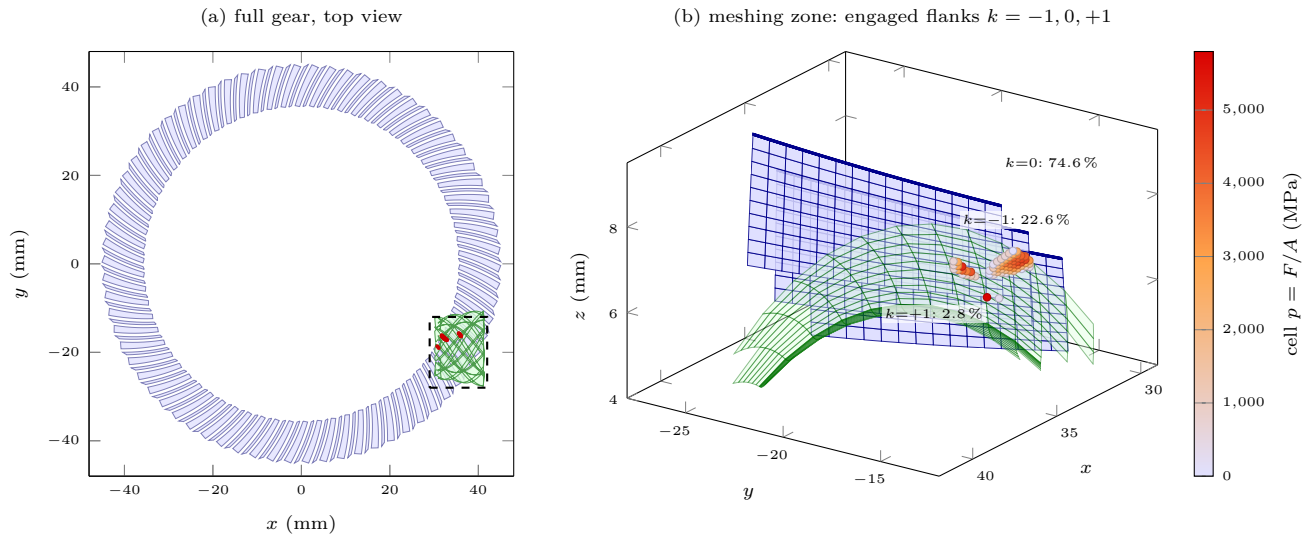
\begin{figure*}[!t]
  \centering
  \begin{tikzpicture}
    \begin{axis}[
      name=ring,
      width=7.2cm, height=7.2cm,
      view={0}{90},
      xlabel={$x$ (\si{\milli\meter})}, ylabel={$y$ (\si{\milli\meter})},
      label style={font=\scriptsize}, ticklabel style={font=\tiny},
      xmin=-48, xmax=48, ymin=-48, ymax=48,
      empty line=jump,
      title={(a) full gear, top view}, title style={font=\scriptsize},
    ]
      \addplot3[draw=blue!45!black!50, fill=blue!9, line width=0.15pt]
        table[x=x, y=y, z=z, col sep=comma] {data/mt_ring_gear.csv};
      \addplot3[draw=green!45!black!70, fill=green!14, fill opacity=0.75,
                line width=0.15pt]
        table[x=x, y=y, z=z, col sep=comma] {data/mt_ring_pinion.csv};
      \addplot3[only marks, mark=*, mark size=0.55pt, red!80!black]
        table[x=x, y=y, z=z, col sep=comma] {data/mt_cloud.csv};
      \addplot3[black, dashed, thick] coordinates
        {(29,-28,7) (42,-28,7) (42,-12,7) (29,-12,7) (29,-28,7)};
    \end{axis}
    \begin{axis}[
      name=zoom,
      at={(ring.east)}, anchor=west, xshift=15mm,
      width=8.6cm, height=7.2cm,
      view={125}{30},
      xlabel={$x$}, ylabel={$y$}, zlabel={$z$ (\si{\milli\meter})},
      label style={font=\scriptsize}, ticklabel style={font=\tiny},
      xmin=29, xmax=42, ymin=-28, ymax=-12, zmin=4, zmax=9.5,
      colormap={pmap}{color(0)=(blue!12) color(0.5)=(orange!70) color(1)=(red!85!black)},
      colorbar, colorbar style={width=2.5mm, ticklabel style={font=\tiny},
        ylabel={cell $p=F/A$ (\si{\mega\pascal})}, ylabel style={font=\scriptsize}},
      point meta min=0, point meta max=5800,
      z buffer=sort,
      title={(b) meshing zone: engaged flanks $k=-1,0,+1$}, title style={font=\scriptsize},
    ]
      \foreach \s in {0,1,2}{
        \addplot3[surf, shader=faceted, mesh/cols=22,
                  colormap={gb}{color(0)=(blue!14) color(1)=(blue!40)},
                  faceted color=blue!55!black, line width=0.05pt, opacity=0.85,
                  forget plot]
          table[x=x, y=y, z=z, col sep=comma] {data/mt_surf_g\s.csv};
      }
      \addplot3[surf, shader=faceted, mesh/cols=22, unbounded coords=jump,
                colormap={pg}{color(0)=(green!12) color(1)=(green!30)},
                faceted color=green!45!black, line width=0.05pt, opacity=0.45,
                forget plot]
        table[x=x, y=y, z=z, col sep=comma] {data/mt_surf_p1.csv};
      \addplot3[scatter, only marks, mark=*, mark size=1.4pt, scatter src=explicit]
        table[x=x, y=y, z=z, meta=p, col sep=comma] {data/mt_cloud.csv};
      \node[font=\tiny, anchor=west, fill=white, fill opacity=0.7, text opacity=1,
            inner sep=1pt] at (axis cs:36.6,-15.6,8.5) {$k{=}{-}1$: \SI{22.6}{\percent}};
      \node[font=\tiny, anchor=west, fill=white, fill opacity=0.7, text opacity=1,
            inner sep=1pt] at (axis cs:32.6,-16.9,8.9) {$k{=}0$: \SI{74.6}{\percent}};
      \node[font=\tiny, anchor=east, fill=white, fill opacity=0.7, text opacity=1,
            inner sep=1pt] at (axis cs:30.4,-19.6,4.6) {$k{=}{+}1$: \SI{2.8}{\percent}};
    \end{axis}
  \end{tikzpicture}
  \caption{Full-wheel multi-tooth loaded contact of case s0003 at
  $T=\SI{64.5}{\newton\meter}$, solved with per-tooth conjugate separation
  fields under one shared rigid rotation, block-diagonal structural
  compliance from a shell Rayleigh--Ritz model of both members, and a
  torque-balance LCP. (a)~All $70$ gear teeth (both flanks, blue) and the
  $7$ pinion teeth in mesh position (green), with the loaded contact cells (red).
  (b)~Zoom with the real flank surfaces: the three engaged gear teeth
  $k=-1,0,+1$ (blue), the mating pinion flank (green), and the loaded cells,
  sharing the torque $74.6\%/22.6\%/2.8\%$; the loaded mesh rotation is
  $124.4''$ versus $111.4''$ for the rigid-tooth patch LCP of
  Fig.~\ref{fig:pressure} ($+12\%$ tooth-bending contribution). This
  pipeline trims boundary-inadmissible edge cells and reports its peak
  through an incomplete-ellipse line-contact formula
  ($\mathrm{CP}_{\max}=\SI{2714}{\mega\pascal}$), so the raw tip-edge spikes
  of Fig.~\ref{fig:pressure} are excluded from the metric by construction;
  individual $F/A$ cells still reach \SI{5.8}{\giga\pascal} at the incipient
  $k{=}{+}1$ contact.}
  \label{fig:multitooth}
\end{figure*}
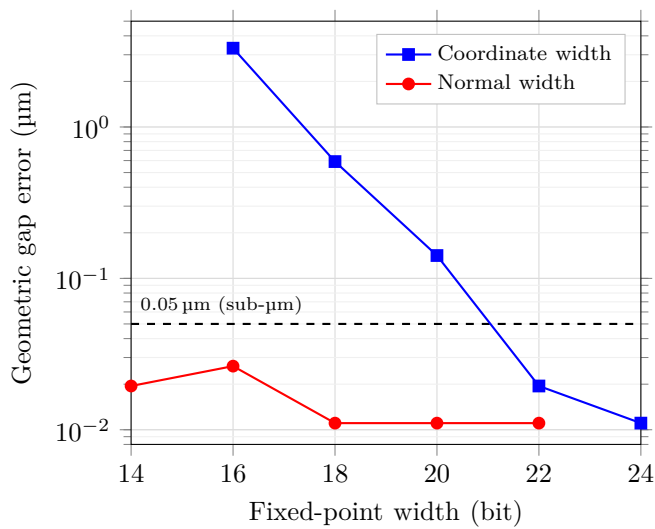
\begin{figure}[!t]
  \centering
  \begin{tikzpicture}
    \begin{axis}[
        width=0.94\columnwidth,
        ymode=log,
        xmin=14, xmax=24,
        ymin=0.008, ymax=5,
        xtick={14,16,18,20,22,24},
        xlabel={Fixed-point width (bit)},
        ylabel={Geometric gap error (\si{\micro\meter})},
        log basis y=10,
        grid=both,
        major grid style={gray!30},
        minor grid style={gray!12},
        tick align=outside,
        legend cell align=left,
        legend style={font=\footnotesize, at={(0.97,0.97)}, anchor=north east, draw=gray!50},
      ]
      \addplot[dashed, thick, black, forget plot]
        coordinates {(14,0.05) (24,0.05)};
      \node[anchor=south west, font=\scriptsize, black]
        at (axis cs:14,0.05) {\SI{0.05}{\micro\meter} (sub-\si{\micro\meter})};
      \addplot[blue, mark=square*, thick, mark options={fill=blue}]
        table[x=coord_bits, y=gap_err_um, col sep=comma]{data/precision_coord.csv};
      \addlegendentry{Coordinate width}
      \addplot[red, mark=*, thick, mark options={fill=red}]
        table[x=norm_bits, y=gap_err_um, col sep=comma]{data/precision_norm.csv};
      \addlegendentry{Normal width}
    \end{axis}
  \end{tikzpicture}
  \caption{Geometric gap error versus fixed-point word length. Coordinate
    word length dominates; a width of $\ge 20$~bit already reaches the
    sub-\si{\micro\meter} regime, so no \texttt{fp64} arithmetic is required.
    Both trends, including the normal-width saturation at $\ge 18$~bit, are
    predicted by the quantization bound of Eq.~\eqref{eq:qbound}.}
  \label{fig:precision}
\end{figure}
\section{Discussion and Limitations}
\label{sec:discussion}

\textbf{Fidelity envelope.}
The physics-direct datapath returns an \emph{engineering-magnitude preview}, not a
certification-grade contact distribution. Its peak pressure tracks the bulk-Hertz
band of the reference LCP solve and reproduces the $W^{1/3}$ scaling exactly, yet
sits $\approx\!22\%$ below the loaded $p_{90}$ ($-13\%$ to $-31\%$ across torque;
Fig.~\ref{fig:accuracy}, Table~\ref{tab:accuracy}) because a single equivalent
Hertzian ellipse cannot resolve the tip- and edge-loading spikes that reach
$\approx\!5\times$ the bulk stress. The loaded transmission error (LTE) behaves the
same way: the preview follows the LTE curve shape and the single-/double-tooth
transition (shared kinematics), but underpredicts its peak-to-peak amplitude by
$19\%$ (contact compliance $0.76\times$; Fig.~\ref{fig:te}). Those absolute peaks,
and any redistribution under misalignment, remain the province of the host
solvers---the Boussinesq+LCP patch \cite{boussinesq1885} and the multi-tooth
shell LTCA whose load sharing is quantified in Fig.~\ref{fig:multitooth}. The preview thus answers, in real time, whether the patch
geometry and stress level are \emph{in family}; it does not replace the LCP for
peak-stress or NVH certification \cite{te2018nvh}.

\textbf{When to use which.}
The three engines are complementary, not competing (Fig.~\ref{fig:nn}).
Physics-direct is the tool for real-time preview, condition monitoring, and HIL:
it extrapolates by Hertzian construction \cite{hamrock1977elliptical}, so on the
LCP-anchored load sweep its out-of-box error ($18.9\%$) undercuts the \emph{trained}
neural surrogate's ($20.3\%$), and on the synthetic domain it avoids the network's
$30\times$ in-box-to-out-of-box error growth. It needs no training set, and every
intermediate---curvature, ellipse axes, pressure---is a physical, fixed-point,
deterministically auditable quantity. A neural surrogate wins only inside a densely
sampled box for bulk offline interpolation, and pays one full LCP per training sample
($\approx\!13$\,min for 400 points) with no extrapolation guarantee. The full LCP
stays the reference whenever the absolute peak distribution is itself the deliverable.
In one line: \emph{physics-direct to watch, neural surrogate to interpolate, LCP to
certify.}

\textbf{Silicon and throughput.}
The four verified SoCs are bit-exact against golden readback but at bake-in /
CSR-BIST scale: they prove the datapath, not sustained throughput. Determinism is
already realized---data-independent, $\sigma=0$ cycle counts
(Table~\ref{tab:latency}, Fig.~\ref{fig:bhseq}(b))---but the board clock is held at
\SI{12.5}{\mega\hertz}. Two open-toolchain constraints drive this. The
\texttt{-nodsp} flow, forced by a nextpnr DSP-cascade placement failure, maps every
multiply into LUTs, inflating area to $33\%$ and yielding $\mathrm{DSP}=0$
(Table~\ref{tab:resource}); and the unpipelined $\mathrm{perp}^2$ gating and the
serialized division/cube-root chains violate setup at the \SI{50}{\mega\hertz}
target. Closing that gap requires pipelined divide/transcendental IP and a
DSP-mapped multiply flow that manages the cascade explicitly. Even then the parallel
gap core is DSP-bound and walls at $L\!\approx\!63$ lanes on this device
(Fig.~\ref{fig:throughput}); wider fan-out needs a larger part or multiplier reuse.

\textbf{Scope beyond hypoid gears.}
Nothing in the datapath is hypoid-specific. Stages A/B assume only a point-sampled
mating surface with unit normals; Stage C assumes the local gap is quadratic over a
fixed stencil; Stage D assumes a counterformal elliptic-Hertz contact. Any
transmission element satisfying those three---spiral-bevel, spur/helical (where the
geometry degenerates gracefully to line-like high-ellipticity contact), cam--follower
pairs, rolling-element raceways---maps onto the same silicon by re-baking the stencil
pseudo-inverse and streaming a different pose schedule; both are host-side CSR
writes, not RTL changes (Fig.~\ref{fig:setup}(b)). The exclusions are equally
explicit: conformal or near-conformal contacts (e.g.\ worm gears, deep-groove
osculation) violate the Hertz half-space premise, and elastohydrodynamic film
effects are outside the model class entirely.

\textbf{Validation scope.}
Quantitative accuracy is anchored on a single flank (s0003). The $W^{1/3}$ pressure
law and the LTE trends should carry over by construction, and the geometry gap is
provably sub-\si{\micro\meter} beyond \SI{20}{bit} coordinates (Fig.~\ref{fig:precision}),
so no \texttt{fp64} path is needed; but multi-flank, multi-alignment, and
wear-modified surfaces remain to be swept before any absolute accuracy figure is
declared portable.

\section{Conclusion}\label{sec:conclusion}

We presented the first silicon-verified fixed-point FPGA pipeline for hypoid
tooth-contact computation, spanning the full geometry--curvature--Hertz chain
(Stages~A/B/C/D) across four distinct SoCs whose JTAG read-back matches a
software golden model bit-for-bit. The design maps to a retired data-center
accelerator (Xilinx~XC7K480T) through an entirely open-source
flow~\cite{openxc7,shah2019yosys} with no floating-point IP; coordinates,
normals, curvatures, and pressure live in audited fixed-point contracts
(Table~\ref{tab:fixedpoint}). A principled partition (Fig.~\ref{fig:partition})
retains reference-frame setup and LCP-grade contact on the host while committing
the per-phase, branch-free contact kernels to fabric.

Our central finding is that a \emph{physics-direct} evaluator is a
viable---and, where it matters, superior---alternative to a neural surrogate for
loaded tooth contact analysis (LTCA). Every intermediate is a physical quantity
(perpendicular-gated gap, closed-form principal curvatures from a baked
pseudoinverse, and a Hamrock--Brewe elliptic Hertz solution), so the pipeline
needs no training data, extrapolates by construction, and is fully auditable.
On an LCP-anchored load sweep the zero-training physics-direct error
(\SI{18.9}{\percent}) undercuts a \emph{trained} MLP (\SI{20.3}{\percent})
outside the fitted box, where the network degrades $\sim$\num{30}$\times$
(Fig.~\ref{fig:nn}).

We report scope honestly: the on-chip result is an engineering-magnitude
\emph{preview}, not a full LCP solve. Preview $p_{\max}$ tracks the body-Hertz
band (mean $\approx$ p90) at \SI{-22}{\percent} versus p90 while preserving the
exact $W^{1/3}$ scaling, and the loaded transmission error (LTE)
peak-to-peak---the dominant NVH excitation---matches within \SI{-19}{\percent}
($1.31''$ vs $1.61''$); the true edge-loaded peak, $\approx\!5\times$ the body
pressure, still requires the host LCP. Timing is exactly
deterministic---fixed cycle counts yield $\sigma=0$ jitter (full
C$\rightarrow$D chain \num{806} cycles; Table~\ref{tab:latency})---and geometry
needs no fp64 (sub-\si{\micro\metre} gap at $\ge\!\num{20}$-bit coordinates).
The gap kernel is DSP-bound, saturating the device near $L=63$ lanes
(Fig.~\ref{fig:throughput}).

Future work retimes the datapath to the \SI{50}{\mega\hertz} streaming target
and validates across multiple tooth flanks and cutter settings. The most
direct payoff is real-time gear \emph{dynamics}: a lumped-parameter drivetrain
model integrates at \si{\micro\second} steps but is conventionally fed by
contact quantities (mesh stiffness, LTE, load-dependent damping) interpolated
from offline tables. With the full chain at \SI{16}{\micro\second} and
$\sigma{=}0$, the preview core can close that loop \emph{live}: each
integration step streams the instantaneous mesh phase and torque over CSR and
receives the load-dependent contact state ($a,b,p_{\max}$, Hertzian
compliance) computed on the actual tooth geometry---no table, no
interpolation error at untabulated loads, and a hard latency bound suitable
for HIL certification of controllers~\cite{fpga2018igbt,fpga2022hil}. A
$L{=}63$-lane build sustains ${\sim}3\times10^{9}$ gap evaluations per
second, enough to re-resolve the mating point within each step of a
\SI{10}{\kilo\hertz} real-time integrator rather than assuming it fixed.

\bibliographystyle{IEEEtran}
\bibliography{references}

\begin{thebibliography}{10}
\providecommand{\url}[1]{#1}
\csname url@samestyle\endcsname
\providecommand{\newblock}{\relax}
\providecommand{\bibinfo}[2]{#2}
\providecommand{\BIBentrySTDinterwordspacing}{\spaceskip=0pt\relax}
\providecommand{\BIBentryALTinterwordstretchfactor}{4}
\providecommand{\BIBentryALTinterwordspacing}{\spaceskip=\fontdimen2\font plus
\BIBentryALTinterwordstretchfactor\fontdimen3\font minus
  \fontdimen4\font\relax}
\providecommand{\BIBforeignlanguage}[2]{{%
\expandafter\ifx\csname l@#1\endcsname\relax
\typeout{** WARNING: IEEEtran.bst: No hyphenation pattern has been}%
\typeout{** loaded for the language `#1'. Using the pattern for}%
\typeout{** the default language instead.}%
\else
\language=\csname l@#1\endcsname
\fi
#2}}
\providecommand{\BIBdecl}{\relax}
\BIBdecl

\bibitem{kolivand2009ease}
M.~Kolivand and A.~Kahraman, ``A load distribution model for hypoid gears using
  ease-off topography and shell theory,'' \emph{Mechanism and Machine Theory},
  vol.~44, no.~10, pp. 1848--1865, 2009.

\bibitem{icm2018bevel}
S.~D. Peng, H.~Ding, G.~Zhang, J.~Y. Tang, and Y.~Tang, ``New determination to
  loaded transmission error of the spiral bevel gear considering multiple
  elastic deformation evaluations under different bearing supports,''
  \emph{Mechanism and Machine Theory}, vol. 137, pp. 37--52, 2019,
  {ScienceDirect} S0094114X18312849.

\bibitem{litvin2004gear}
F.~L. Litvin and A.~Fuentes, \emph{Gear Geometry and Applied Theory},
  2nd~ed.\hskip 1em plus 0.5em minus 0.4em\relax Cambridge University Press,
  2004.

\bibitem{te2023surrogate}
M.~Willecke, J.~Brimmers, and C.~Brecher, ``Surrogate model based prediction of
  transmission error characteristics based on generalized topography
  deviations,'' \emph{Forschung im Ingenieurwesen}, vol.~87, pp. 431--440,
  2023.

\bibitem{ml2025hertzian}
F.~Bruzzone, D.~Fabbri, and C.~Rosso, ``Machine learning surrogate models for
  {Hertzian} contact stress prediction in gear design: A comparative study of
  multiple approaches,'' \emph{Results in Engineering}, 2025, {ScienceDirect}
  S3050475925008073.

\bibitem{pinn2024contact}
T.~Sahin, M.~von Danwitz, and A.~Popp, ``Solving forward and inverse problems
  of contact mechanics using physics-informed neural networks,'' \emph{Advanced
  Modeling and Simulation in Engineering Sciences}, vol.~11, 2024, {DOI}
  10.1186/s40323-024-00265-3.

\bibitem{pinn2025energy}
J.~Bai, Z.~Lin, Y.~Wang, J.~Wen, Y.~Liu, T.~Rabczuk, Y.~Gu, and X.-Q. Feng,
  ``Energy-based physics-informed neural network for frictionless contact
  problems under large deformation,'' \emph{Computer Methods in Applied
  Mechanics and Engineering}, 2025, {arXiv}:2411.03671.

\bibitem{aimbs2025gear}
M.~Willecke, J.~Brimmers, and C.~Brecher, ``Accelerating {FE}-based gear mesh
  calculations in dynamic multi-body simulations with {AI},'' \emph{Forschung
  im Ingenieurwesen (Engineering Research)}, 2025, {DOI}
  10.1007/s10010-025-00783-5.

\bibitem{fpga2018igbt}
C.~Liu, R.~Ma, H.~Bai, Z.~Li, F.~Gechter, and F.~Gao, ``{FPGA}-based real-time
  simulation of high-power electronic system with nonlinear {IGBT}
  characteristics,'' \emph{IEEE Journal of Emerging and Selected Topics in
  Power Electronics}, vol.~7, no.~1, pp. 41--51, 2019, {DOI}
  10.1109/JESTPE.2018.2873157.

\bibitem{fpga2020latency}
C.~Liu, H.~Bai, S.~Zhuo, X.~Zhang, R.~Ma, and F.~Gao, ``A latency-insensitive
  design approach to programmable {FPGA}-based real-time simulators,''
  \emph{Electronics}, vol.~9, no.~11, p. 1838, 2020.

\bibitem{fpga2022hil}
M.~Sotero, G.~Fontenele, F.~Dicler, M.~Neves, L.~F. Corr\^ea, and M.~Aredes,
  ``An {FPGA}-based hardware-in-the-loop implementation of power electronics
  circuits using a generic real-time simulator,'' in \emph{Brazilian Power
  Electronics Conference (COBEP)}, Jo\~ao Pessoa, Brazil, 2021, pp. 1--8, {DOI}
  10.1109/COBEP53665.2021.9684098.

\bibitem{hamrock1977elliptical}
B.~J. Hamrock and D.~E. Brewe, ``Simplified solution for elliptical-contact
  deformation between two elastic solids,'' \emph{ASME Journal of Lubrication
  Technology}, vol.~99, no.~4, pp. 485--487, 1977.

\bibitem{shah2019yosys}
D.~Shah, E.~Hung, C.~Wolf, S.~Bazanski, D.~Gisselquist, and M.~Milanovi\'c,
  ``{Yosys+nextpnr}: An open source framework from {Verilog} to bitstream for
  commercial {FPGAs},'' in \emph{IEEE Int. Symp. Field-Programmable Custom
  Computing Machines (FCCM)}, 2019, {arXiv}:1903.10407.

\bibitem{openxc7}
{openXC7 contributors}, ``{openXC7}: Open-source {FPGA} toolchain for
  {AMD/Xilinx} 7-series (yosys + nextpnr-xilinx + prjxray),''
  \url{https://github.com/openXC7}, 2024.

\bibitem{te2018nvh}
A.~Palermo, L.~Britte, K.~Janssens, D.~Mundo, and W.~Desmet, ``The measurement
  of gear transmission error as an {NVH} indicator: Theoretical discussion and
  industrial application via low-cost digital encoders to an all-electric
  vehicle gearbox,'' \emph{Mechanical Systems and Signal Processing}, vol. 110,
  pp. 368--389, 2019, {ScienceDirect} S0888327018301249.

\bibitem{evgear2025nvh}
K.~Horvath and D.~Feszty, ``Surface waviness of {EV} gears and {NVH} effects
  — a comprehensive review,'' \emph{World Electric Vehicle Journal}, vol.~16,
  no.~9, p. 540, 2025.

\bibitem{efficient2025hypoid}
K.~Rong, J.~Tang, Z.~Tian, B.~Song, H.~Li, and H.~Ding, ``A novel
  accurate-efficient loaded contact analysis method for hypoid gears based on
  ease-off topography discretization and {TE}-interference assessment,''
  \emph{Mechanism and Machine Theory}, vol. 209, 2025, {ScienceDirect}
  S0094114X25001016.

\bibitem{semianalytical2024bevel}
Y.~Liu, L.~Chen, X.~Mao, and D.~Shangguan, ``A semi-analytical loaded contact
  model and load tooth contact analysis approach of ease-off spiral bevel
  gears,'' \emph{Machines}, vol.~12, no.~9, p. 623, 2024.

\bibitem{multitooth2025hypoid}
J.~Pang, S.~Liu, C.~Song, and C.~Liang, ``General multi-tooth contact analysis
  of spiral bevel and hypoid gears with arbitrary shaft angles considering the
  point clouds reconstruction of gear surface based on deep neural network,''
  \emph{Mechanism and Machine Theory}, vol. 214, p. 106139, 2025, {DOI}
  10.1016/j.mechmachtheory.2025.106139.

\bibitem{robust2024hypoid}
X.~Wei, Y.~Wang, W.~Zhang, and T.~C. Lim, ``Robust optimization of hypoid gear
  contact performance considering tooth form error: Design sensitivity and
  {Pareto} front,'' \emph{Mechanism and Machine Theory}, 2024, {DOI}
  10.1016/j.mechmachtheory.2024.105754.

\bibitem{nie2024hypoid}
S.~Nie, J.~Chen, and S.~Liu, ``Research on noise reduction of drive axle hypoid
  gear based on tooth surface mismatch modification,'' \emph{Advances in
  Mechanical Engineering}, 2024, {DOI} 10.1177/16878132241228195.

\bibitem{fpga2026optctrl}
T.~Desai, B.~Plancher, and R.~I. Bahar, ``Real-time, energy-efficient,
  sampling-based optimal control via {FPGA} acceleration,'' {arXiv}:2601.17231,
  2026.

\bibitem{fpgadt2024reduction}
M.~Ciklamini and M.~Cejnek, ``Enhancing digital twin performance through
  optimizing graph reduction of finite element models,'' \emph{Scientific
  Reports}, vol.~15, p. 37777, 2025, {DOI} 10.1038/s41598-025-20571-z.

\bibitem{fpga2007raytriangle}
T.~Kim and B.~Nam, ``Fast ray-triangle intersection computation using
  reconfigurable hardware,'' \emph{Lecture Notes in Computer Science
  (Springer)}, 2007.

\bibitem{collision2004hardware}
G.~Knittel and G.~Zachmann, ``High-performance collision detection hardware,''
  University of Bonn, Informatik II, Tech. Rep. CG-2003-3, 2004.

\bibitem{fpga2016motionplan}
S.~Murray, W.~Floyd-Jones, Q.~Ying, D.~J. Sorin, and G.~Konidaris, ``Robot
  motion planning on a chip,'' in \emph{Robotics: Science and Systems (RSS)},
  2016.

\bibitem{boussinesq1885}
J.~Boussinesq, \emph{Application des potentiels \`a l'\'etude de l'\'equilibre
  et du mouvement des solides \'elastiques}.\hskip 1em plus 0.5em minus
  0.4em\relax Paris: Gauthier-Villars, 1885, {Half}-space influence function
  underlying the reference LTCA.

\bibitem{johnson1985contact}
K.~L. Johnson, \emph{Contact Mechanics}.\hskip 1em plus 0.5em minus 0.4em\relax
  Cambridge University Press, 1985.

\bibitem{savitzky1964smoothing}
A.~Savitzky and M.~J.~E. Golay, ``Smoothing and differentiation of data by
  simplified least squares procedures,'' \emph{Analytical Chemistry}, vol.~36,
  no.~8, pp. 1627--1639, 1964, {Stage} C's baked pseudo-inverse MAC is the
  two-dimensional analogue of these precomputed least-squares convolution
  weights.

\end{thebibliography}

\end{document}